\documentclass[12pt]{article}

\usepackage[english]{babel}
\usepackage[usenames]{color}
\usepackage[cp1250]{inputenc}
\usepackage{amsfonts}
\usepackage{amsthm}
\usepackage{graphicx}
\usepackage{epsfig}

\theoremstyle{plain}
\newtheorem{df}{Definition}

\newtheorem{tw}[df]{Theorem}
\newtheorem{lem}[df]{Lemma}

\newtheorem{sps}[df]{Observation}

\newtheorem{zal}[df]{Assumption}

\begin{document}
\newcommand{\bea}{\begin{eqnarray}}
\newcommand{\eea}{\end{eqnarray}}
\newcommand{\be}{\begin{equation}}
\newcommand{\ee}{\end{equation}}
\newcommand{\beas}{\begin{eqnarray*}}
\newcommand{\eeas}{\end{eqnarray*}}
\newcommand{\bs}{\backslash}
\newcommand{\bc}{\begin{center}}
\newcommand{\ec}{\end{center}}

\title{Optimal encoding on discrete lattice\\with translational invariant constrains\\
using statistical algorithms.}

\author{Jarek Duda}

\date{\it \footnotesize Jagiellonian University, Cracow Poland, \\
\textit{email:} dudaj@interia.pl}

\maketitle

\begin{abstract}
In this paper will be presented methodology of encoding
information in valuations of discrete lattice with some
translational invariant constrains in asymptotically optimal way.
The method is based on finding statistical description of such
valuations and changing it into statistical algorithm, which
allows to construct deterministically valuation with given
statistics. Optimal statistics allow to generate valuations with
uniform distribution - we get maximum information capacity this
way. It will be shown that we can reach the optimum for
one-dimensional models using maximal entropy random walk and that
for the general case we can practically get as close to the
capacity of the model as we want (found numerically: lost
$10^{-10}$bit/node for Hard Square). There will be also presented
simpler alternative to arithmetic coding method which can be used
as cryptosystem and data correction method too.
\end{abstract}

\section{Introduction}
Consider all projections $\mathbb{Z}^2\to\{0,1\}$. In this way we
can store 1 bit/node (point of the space). Now introduce some
constrains, for example: there cannot be two neighboring "1" (each
node has 4 neighbors) - it's so called Hard Square model(HS). It
will occur that this restriction reduces the informational
capacity to $H_{HS}\cong 0.5878911617753406$ bits/node.

The goal of this paper is to introduce methodology of encoding
information in such models as near their capacity as required.\\

We will call a \emph{model} such triplet - space ($\mathbb{Z}^2$),
alphabet ($\{0,1\}$) and some constrains. It's \emph{elements} are
all projections fulfilling the constrains - we can think about
them as \emph{valuations} of nodes. Now the number of all such
valuations over some finite set of nodes($A$) will asymptotically
grow exponentially $N\cong 2^{\#A H}$. Because in the possibility
of choosing one of $N$ choices can be stored $\lg(N)$ bits, this
$H$ (Shannon's entropy) is the maximal capacity in bits/node we
can achieve.

We can really store $\lg(N)$ bits in choosing one of $N$ choices,
only if all of them are equally probable only. So to get the whole
available capacity, we have to make that all possible valuations
are equally probable. Unfortunately the space of valuations over
infinite space is usually quite complicated. But thanks of
translational symmetry, elements should have the same local
statistical behavior. If we find it and valuate the space
accordingly, we should get near to the uniform distribution over
all elements. The statistical algorithm have to encode
some information in generating some specific valuation,
fulfilling the optimal statistics of the space.\\

Statistical description ($p$) is a function, which for every
finite set (\emph{shape}) gives the probability distribution of
valuations on it (\emph{patterns}). Thanks of the translational
invariance, we can for example write $p(01)$ - the probability
that while taking any two neighboring nodes, they give '01'
pattern. In one dimension we can find the optimal statistical
description using pure combinatorics. In higher it's much more
complicated, but we can for example divide the space into short
stripes, create new alphabet from their valuations and just use
the one-dimensional method.

Having the statistical description, we can use it to construct the
statistical algorithm. For example divide the space into straps
and valuate them succeedingly. Now for succeeding nodes, depending
on the valuations of already valuated neighbors, we get some
probability from created previously table. According to this
probability we valuate the node, encoding some information.\\

\textbf{Examples of usage}: We can think about for example hard
disk, locally as valuating nodes (let say - magnetizing round
dots) of two-dimensional lattice with 0 or 1 (without constrains).\\
\begin{figure}[h]
    \centering
        \includegraphics[width=5cm]{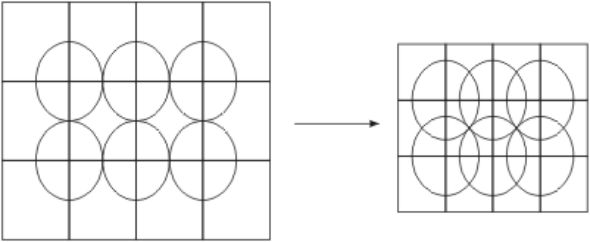}
        \caption{Rescaling the lattice without changing the size of magnetic dot.}
\end{figure}

Now let us change the lattice constant, as on fig. 1 - we have
$\sqrt{2}^2=2$ times more nodes, but we get some constrains - like
in HS - so the capacity is now: $2*0.587\cong 1.17$ - we get 17\%
greater capacity. We've got it by more precise positioning of the
head - it's technically easier to achieve than shrinking the dot.
We
will see that going further, we can increase the capacity potentially to infinity.\\

We can use statistical algorithm approach also to generate random
states (e.g. spin alignment) in statistical physics. Usually we
use Monte-Carlo methods - to generate "good" alignment we have to
make many runs (the more, the better).

But using for example many of such alignments, we can approximate
its (local) statistical description with assumed "goodness". Now
using statistical algorithm, we can generate so "good" alignments in one run.\\

In the \textbf{second section} we will see how to solve
analytically the one-dimensional model - find its optimal
description and capacity. We will motivate that it should be
Shannon's entropy. To find the optimal description we will just
average over all elements. In this case the statistical algorithm
will be just Markov process - we will valuate node by node from
left to right and the probability distribution for a node is found
using only the valuation of the previous one. We get this way
random walk on a graph (of symbols), which maximizes global
entropy (\cite{merw}). This approach can be generalized to other
than uniform distributions of sequences, by introducing some
vertex/edge potentials.

In the \textbf{third section} there will be presented asymmetric
numeral systems - a generalization of numeral systems, which are
optimized for encoding sequences of equiprobable digits into which
the probability distribution of digits is given. It's natural way
to encode data using given statistical algorithm. This algorithm
can be alternative for widely used arithmetic coding method: in
one table check it compress/decompress a few bits (a symbol) and
have option that the output is encrypted, probably very well. It
has also very nice data correction properties.

In the \textbf{fourth section} there will be introduced formality
for general models. It will be shown that for "reasonable" models:
$X=\mathbb{Z}^n$, translative invariant constrains with finite
range and which are "simple" - valuation of some nodes cannot
enforce valuation of distant enough ones - we can speak about
their entropy, which is positive.

In the \textbf{fifth section} we will introduce methodology of
statistical description. Now we will define the optimal
description straightforward as the average over all elements.
Unfortunately, in spite of giving many arguments, I couldn't prove
existence of such average - we will assume it and see that it's
equivalent to vanishing of long-range correlation.  There will be
also introduced alternative definition of optimality (LOC) - in a
finite set of nodes, separated topologically from the rest of the
space, all valuations are equally probable. Then there will be
discussed how to generate elements for given statistical
description ('almost' statistical algorithm): fix some order for
nodes and get probability distribution for a node using valuation
of the previous ones.

In the \textbf{sixth section} we will analyze some algorithms as
above, but this time there can be infinite number of previous
nodes. We will assume that the probability can be determined only
by valuation of neighboring nodes. We will analyze two simple
types of algorithms: valuate separately subsets inside which
constrains doesn't work or just take a random order. We will get
near optimum this way and explain why we can't get it in this way.
There will be also described how to iteratively generate
approximations of uniform distribution of elements on a finite
set. The longer we will iterate the process, the nearer uniform
probability we get. We can use this generator to find numerically
approximation of the optimal description.

In the \textbf{seventh section} it will be finally shown how to
get as close to the optimal algorithm as we want. We will do it by
narrowing the space, so that it has only one infinite dimension
and divide it into "straps", for which we can use one-dimensional
analytical methods. There will be shown numerical results, showing
that we are really tending quickly to the optimum in this way.

\section{One-dimensional model}
In this section we will look at the following transitively
invariant model:
\begin{df} One-dimensional model \emph{will be called a triplet:}
$(X,\mathcal{A},M)$:\\
space $X=\mathbb{Z}$\\
alphabet $\mathcal{A} $\emph{  - finite set of symbols  } \\
constrains $M:\mathcal{A}^2\to\{0,1\}$\\
\\\emph{Now} elements \emph{of this model are} $V(X)$, \emph{where for} $A\subset X$:
\begin{equation}V(A):=\{v:A\to\mathcal{A}\vert\forall_{i\in
A\cap(A-1)}M(v_i,v_{i+1})=1\}
\end{equation}
\end{df}
\subsection{Blocking symbols to reduce constrains to range one}
I will shortly justify, that any general one-dimensional,
translatively invariant model, can be easily reduced to above
(with constrains of range one):

 Let $l$ be the largest range of constrains, for example
$v_k=a\Rightarrow v_{k+l}=b$. Take $\mathcal{A}^l$ as the new
alphabet, grouping $l$ consecutive symbols.

Now we can construct matrix as above:
\begin{eqnarray*}&M_{(v_i)_{i=1..l}(w_i)_{i=1..l}}=1\Leftrightarrow\\\Leftrightarrow&\left(\forall_{i=2..l}\ v_i=w_{i-1}
\textrm{ and the sequence } v_1 v_2 ... v_l w_l \textrm{ is
consistent with the constrains}\right)\end{eqnarray*}
So we can restrict to the model defined above ($l=1$).\\

Let's visualize it to analyze some \textbf{example}:
\begin{df}$k$-model:\\
$X=\mathbb{Z}\quad A=\{0,1\}\quad$ \emph{constrain}: after $1$
follows at least $k$ zeros.
\end{df}
Because constrains have range $k$, we should group $k$ symbols
into one of new symbols "0","1", ..."k":
\begin{itemize}
\item "0" : there were at least $k$ zeros before (there can be $1$ now)
\item "$i$" : $k-i+1$ positions before was the last $1$ (in $i$ positions
there can be $1$)
\end{itemize}
In states different than "0", we have to put $0$.

 So the whole algorithm (Markov process) is defined by the probability of putting
$1$ while in state "0" - denote it $q$ ($\tilde{q}:=1-q$). Denote $p_i$ - the probability of state $i$.\\
Make one step ($p:=p_0$):
$\left\{\begin{array}{l}p_k=pq\\p_i=p_{i+1}\quad\hbox{for
}i\in\{1,...,k-1\}\\p=p_1+p\tilde{q}\end{array}\right.$\\
So $p_1=p_2=...=p_k=pq$\\
$p=1-p_1-p_2-...-p_k=1-kpq\qquad\quad$${p=\frac{1}{1+kq}}$\\

We will explain later, that in a symbol with probability
distribution $q/\tilde{q}$ is stored at average
$h(q):=-q\lg(q)-\tilde{q}\lg(\tilde{q})$ bits. So the entropy of
this model is the maximum over all algorithms: \be
H=\max_{q\in[0,1]}H_q=\max_{q\in[0,1]}\frac{h(q)}{1+kq}.\ee

In the introduction we've seen the example that this method can be
used to store more data on two-dimensional plane. We've just found
analytic expression for the one-dimensional case - we have
constant length of "magnetic dot" - say $d$, but we don't assume
that potential positions cannot intersect (if we assume that - we
can store 1bit/length $d$). We assume only that two $1$
(magnetized dot) cannot intersect.

Let say that we can position the dot with precision
$\frac{d}{k+1}$. That means exactly that after $1$ there have to
be $k$ zeros - analyzed model. We can now easily count that using
$k+1$ times more precise positioning, we get:\\

\begin{figure}[h]
    \centering
        \includegraphics{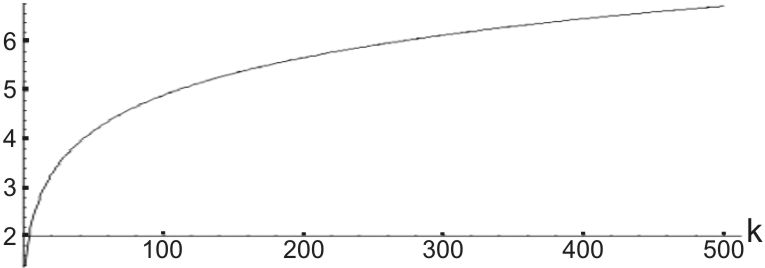}
        \caption{Informational capacity in bits/'old node' for rescaled $k$ - models}
        \label{f1}
\end{figure}

\begin{tabular}{|c|c|c|c|c|c|c|c|c|c|c|c|c|c|}
\hline $k$&0&1&2&3&4&5&6&7&8&9&10&11&12\\
\hline
benefit(\%)&0&39&65&86&103&117&129&141&151&160&168&176&183\\
\hline \end{tabular}\\

more capacity of information. For larger $k$ we get fig. \ref{f1}
It goes slowly to infinity with $k\to\infty$ - we could increase
the capacity of information potentially as much as we need by more
precise head positioning.
\subsection{Maximal entropy random walk}
Let's think how much information can be stored in such
sequence/path of length $k$?\\
Denote $N_k=\#V\left(\,\overline{k}\,\right)\ $
$\left(\,\overline{k}:=\{0,1,...,k-1\}\right)$ the number of such
sequences of length $k$. We can store some information by choosing
one of them. To choose one of $2^n$ elements we need $n$ bits. So
in sequence of length $k$ we can store about $\lg(N_k)$ bits. The
average number of bits we can store in one node is:
$H:=\lim_{k\to\infty}H^k$
\begin{equation} H^k:=\frac{lg \left(\#V\left(\,\overline{k}\,\right)\right)}{k} \end{equation}
Let's introduce vector $c^k:=(c^k_a)_{a\in\mathcal{A}}$, where
 $c^k_a:=\#\{\gamma\in V\left(\,\overline{k}\,\right):\gamma_{k-1}=a\}$.\\
Sequences of length $k+1$ are obtained from sequences of length
$k$: \ \be c^{k+1}=c^k \cdot M\ee
\begin{tw}[Frobenius - Perron] Irreducible matrix with real,
nonnegative coefficients have dominant real, nonnegative eigenvalue.
Corresponding eigenvector has real, positive coefficients.
\end{tw}
In our case reducibility would mean that starting from proper
subset of alphabet we have to stay in it - we could decompose it
into irreducible cases.

Assuming irreducibility, we can say: \be H=lg\lambda \ee where $\lambda$ is the dominant eigenvalue of $M$. \\

Look at the normalized corresponding eigenvector: $\phi^TM=\lambda
\phi^T,\ \sum_{a\in\mathcal{A}}\phi_a=1$.\\
We could think that it is probability distribution of symbols in
elements...

It's not true: it's the distribution of symbols on the end of a
sequence which is unbounded in one direction.

We can use this probability distribution to initiate algorithm we
will find.\\

Basing on above derivation we will find the probability
distribution inside such sequences - unbounded in both directions.
Now we have to expand in both sides.

 For some
$(v_i)_{i=0..m-1},\ k\in \mathbb{N},\ a,b\in \mathcal{A}$,
consider all paths from $a$ to $b$ of length $2k+m$, which has $v$
in the middle:
$$\Gamma_v^{k,a,b}:=\{\gamma:\overline{m+2k}\to \mathcal{A},\ \gamma_k
\gamma_{k+1}...\gamma_{k+m-1}=v, \ \gamma_0=a,\
\gamma_{2k+m-1}=b\}$$

We will call $v$ \emph{a pattern}, its domain ($\overline{m}$) -
its \emph{shape} and the extremal nodes of the domain - its
\emph{boundary}: $\quad\grave{v}:=v_0,\ \acute{v}:=v_{m-1}$.

We want to count allowed paths among them. Define matrix $C_v^k$:
\begin{equation} (C_v^k)_{a,b}:=\#\{\gamma\in \Gamma_v^{k,a,b}\cap V\left(\,\overline{m+2k}\,\right)\}=
\sum_{\gamma\in \Gamma_v^{k,a,b}}M_{\gamma_0 \gamma_1}M_{\gamma_1
\gamma_2}...M_{\gamma_{m+2k-2} \gamma_{m+2k-1}} \label{cv}
\end{equation}
The second form uses that such multiplication is equal 1 for
allowed sequences and 0 otherwise. This form suggests
generalization to other matrices and will be considered later. It
also suggests $k\to k+1$ iteration:
$$\qquad C_v^{k+1}=M\cdot C_v^k
\cdot M$$

For the dominant eigenvalue we have left/right eigenvectors:
\begin{equation} \phi^T M=\lambda \phi^T\qquad\qquad M\psi=\lambda \psi\end{equation}
This time we take normalization $\phi^T \psi=1$. \\Usually we will
consider real symmetric matrices, like for undirected graphs, for
which $\phi=\psi$. Generally if $\phi$ is not real, we should use
$\overline{\phi^T}$ instead of $\phi^T$.

$M$ has unique dominant eigenvalue, so for large $k$ we can
assume:
\begin{equation} M^k\cong \lambda^k \psi\phi^T\end{equation}

This asymptotic behavior is the key point: it allows us to average
over infinitely long sequences. Presented approach can be
generalized to degenerated dominant eigenvalue by projecting into
its eigenspace, but only if all dominant eigenvalues: with the
same absolute value has also the same phase. In other case the
final value would depend on $k$.

Substituting we get
\begin{equation}C_v^k\cong \lambda^{2k} \psi \phi^T C_v^0 \psi \phi^T=\lambda^{2k}(\psi\phi^T) \phi_{\grave{v}} \psi_{\acute{v}}
\end{equation}
because $(C_v^0)_{a,b}=1$ if $a=\grave{v},\ b=\acute{v}$ and 0
otherwise.

Let's look at the probability distribution of allowed patterns on
the given shape ($\overline{m}$) inside such infinite sequences
($k\to\infty$).
\begin{equation}
p_v=\lim_{k\to\infty}\frac{\sum_{a,b\in\mathcal{A}}(C^k_v)_{a,b}}
{\sum_{w\in
V(\overline{m})}\sum_{a,b\in\mathcal{A}}(C^k_w)_{a,b}}=
\frac{\phi_{\grave{v}}\psi_{\acute{v}}}{\sum_{w\in
V(\overline{m})}\phi_{\grave{w}}\psi_{\acute{w}}}
\end{equation}
Notice that we can forget about the normalization assumption in
this equation.\\

We get
\begin{enumerate}
\item{Patterns on the same shape and with equal boundary values are equally probable.
In other words - after fixing values on the boundary of some set,
probability distribution of valuations inside is uniform. We will
see later that it can be generalized into higher dimensions.}
\item{For $m=1$ we get the probability distribution of symbols:
\begin{equation} p_a=\frac{\phi_a \psi_a}{\sum_{b\in \mathcal{A}} \phi_b
\psi_b}=\frac{\phi_a \psi_a}{\phi^T\psi}
\end{equation}}
\item{For $m=2$ we get
\begin{equation} p_{ab}=\frac{\phi_a M_{ab}\psi_b}{\sum_{a',b'\in \mathcal{A}}
\phi_{a'} M_{a'b'} \psi_{b'}}=\frac{\phi_a \psi_b
M_{ab}}{\phi^TM\psi}= \frac{\phi_a \psi_b M_{ab}}{\lambda
\phi^T\psi}
\end{equation}}
\item The probability that from vertex $a$ we will jump to vertex
$b$ is
\begin{equation}
S_{ab}=\frac{p_{ab}}{p_a}=\frac{M_{ab}}{\lambda}\frac{\psi_b}{\psi_a}
\end{equation}
\end{enumerate}
The Markov process defined this way ($P(a\rightarrow b)=S_{ab}$)
reproduces original statistics of the space of infinite allowed
sequences with uniform probability distribution. In fact it gives
uniform probability among finite paths also: for any two points,
each allowed path of given length($k$) between them has the same
probability
\begin{equation}
P(\textrm{path
}\gamma_0\gamma_1..\gamma_k)=\frac{1}{\lambda^k}\frac{\psi_{\gamma_k}}{\psi_{\gamma_0}}\end{equation}
Taking $S^k$ or multiplying above by the combinatorial factor, we
get:
\begin{equation}(S^k)_{ab}=\frac{(M^k)_{ab}}{\lambda^k}\frac{\psi_b}{\psi_a}.\label{sk}\end{equation}

It suggest the statistical algorithm: after symbol $a$ choose the
next one with $(S_{ab})_b$ probability distribution.\\
In this way we get uniform distribution among sequences - get
maximal entropy.

We will also calculate, that it gives
maximal possible entropy $\lg(\lambda)$.\\

Firstly look at the problem: we have a sequence of $0$ and $1$ in
which the probability of $1$ is fixed to some $p\in(0,1)$. Let's
calculate how much information corresponds asymptotically to one
digit. Denote $\tilde{p}:=1-p$
\begin{eqnarray*}
{n \choose pn}&=& \frac{n!}{(pn)!(\tilde{p}n)!}\cong
(2\pi)^{-1/2}\frac{n^{n+1/2}e^n}{(pn)^{pn+1/2}(\tilde{p}n)^{\tilde{p}n+1/2}e^n}=\\&=&
(2\pi np\tilde{p})^{-1/2}p^{-pn}\tilde{p}^{-\tilde{p}n}=(2\pi
np\tilde{p})^{-1/2}2^{-n(p\lg p+\tilde{p} \lg{\tilde{p}})}
\end{eqnarray*}
\begin{equation} H_p=\lim_{n\to\infty} \frac{\lg{n \choose
{pn}}}{n}=-p\lg p-\tilde{p} \lg{\tilde{p}}=:h(p)
\end{equation}
where we've used the Stirling's formula:
$\lim_{n\to\infty}\frac{n!}{\sqrt{2\pi
n}\left(\frac{n}{e}\right)^n}=1$

\begin{figure}[h]
    \centering
        \includegraphics[width=5.5cm]{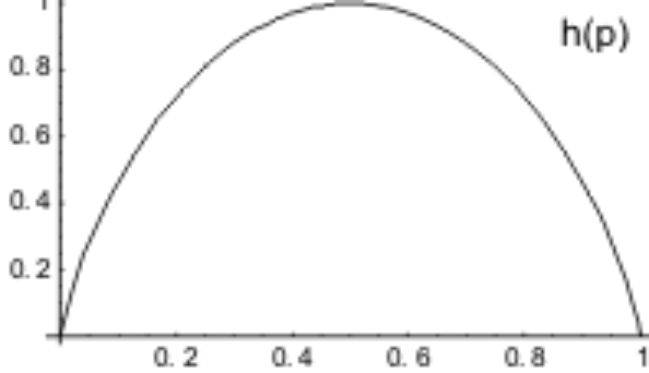}
        \caption{Entropy for fixed $p$.}
\end{figure}

That means that when we choose from all sequences, which number
grows like $2^n$, these with given asymptotical probability of $1$
($p$), we get $2^{nh(p)}$ sequences. If we restrict to sequences
with uncorrelated bits with $p=1/2$, we would get every sequence
with the same probability - the uniform distribution of sequences.
While storing information in this distribution we get the maximum
capacity: 1bit/digit.

Inserting some correlations, favoring some symbols, or part of the
space would create redundancy.

Now when we have larger alphabet $\mathcal{A}$ and a probability
distribution \\$p\ :\ \sum_{a\in\mathcal{A}}\ p_a=1$, the mean
number of information per symbol is (by induction)
\begin{equation}
H_p=-\sum_{a\in\mathcal{A}}p_a \lg{p_a}
\end{equation}
it's the Shannon's entropy - in a symbol with given
probability distribution $p$ we can store $H_p$ bits.\\

In the case of found above Markov process,
$S=(S_{ab})_{a,b\in\mathcal{A}}$:
\begin{enumerate}
\item{$\forall_{a,b\in\mathcal{A}}\ M_{ab}\geq S_{ab}\geq 0$}
\item{$\forall_{a\in\mathcal{A}}\sum_{b\in\mathcal{A}}S_{ab}=1$}
\item{$pS=p,\ \sum_{a\in\mathcal{A}}p_a=1$}
\end{enumerate}

Generating path for a Markov process is a sequence of independent
choices - the entropy of choosing one of sequences is the sum of
entropies for single choices. So the average amount of information
per symbol is:
\begin{eqnarray*}
H&=&-\sum_a p_a \sum_b S_{ab}\lg{S_{ab}}=\\&=&-\sum_a \frac{\phi_a
\psi_a}{\phi^T\psi} \sum_b
\frac{M_{ab}}{\lambda}\frac{\psi_b}{\psi_a}\lg{\frac{1}{\lambda}\frac{\psi_b}{\psi_a}}=
\frac{-1}{\lambda \phi^T\psi}\sum_{a,b}\phi_a M_{ab} \psi_b
\lg{\frac{1}{\lambda}\frac{\psi_b}{\psi_a}}=\\&=&\frac{\phi^TM\psi\lg{\lambda}}{\lambda
\phi^T\psi}+\frac{1}{\lambda \phi^T\psi}\sum_{a,b}(\phi_a
(\lg{\psi_a}) M_{ab}\psi_b-\phi_a
M_{ab}\psi_b(\lg{\psi_b}))=\\&=&\lg{\lambda}+\frac{1}{\lambda
\phi^T\psi}\sum_{a,b}(\phi_a (\lg{\psi_a}) \lambda \psi_b-\phi_a
\lambda \psi_b(\lg{\psi_b}))=\lg{\lambda}
\end{eqnarray*}
We know that this is the limit for all allowed sequences, so among
stochastic processes consistent with the graph ($S_{ab}\leq
M_{ab}$) we cannot get higher entropy - encoding information
 this way gives the maximum capacity.\\

\subsection{Generalization to other distributions of sequences}
We've focused on 0/1 matrices, but above derivations are more
general. Analogously we can enforce that paths are not equally
probable, but their probability is proportional to the sum
(integral) of some potential along the way: $$P(\textrm{path}\
\gamma_0 \gamma_1...\gamma_k)\ \textrm{is proportional to}\
e^{-(\frac{1}{2}V(\gamma_0)+V'(\gamma_0,\gamma_1)+V(\gamma_1)+...+
V'(\gamma_{k-1},\gamma_k)+\frac{1}{2}V(\gamma_k))}$$ where $V,\
V'$ are some freely chosen real vertex/edge potentials.

In discretized euclidean path integrals, the potential says about
the probability that the particle will decay in given place (for
example by going to an absorbing vertex). In the presented
approach, the particle doesn't decay. The potential
says about the probability that the particle will use given vertex/edge.\\

Looking at (\ref{cv}) we see that to achieve this potential, we
should choose: \be
M_{ij}=e^{-\left(\frac{1}{2}V(i)+V'(i,j)+\frac{1}{2}V(j)\right)}
\ee So if we want to use only vertex (edge) potential, we can set
$V'=0$ ($V=0$). If $V'$ is symmetric, $\psi=\phi$. To forbid some
verexes/edges (like before), we can set their potential to
infinity. We can also choose nonzero diagonal elements to allow
self-loops.

Now for a connected weighted graph Frobenius - Perron theorem
still works: we get the dominant eigenvalue ($\lambda$) and
corresponding normalized right eigenvector $\psi>0,\
\sum_i\psi_i^2=1$ and as before: \be P(\textrm{path
}\gamma_0\gamma_1,..,\gamma_k)=\frac{e^{-(\frac{1}{2}V(\gamma_0)+V'(\gamma_0,\gamma_1)+V(\gamma_1)+...+
V'(\gamma_{k-1},\gamma_k)+\frac{1}{2}V(\gamma_k))}}{\lambda^k}\frac{\psi_{\gamma_k}}{\psi_{\gamma_0}}\ee

\be p_i=\psi_i^2\qquad\qquad
(S^k)_{ab}=\frac{(M^k)_{ab}}{\lambda^k}\frac{\psi_b}{\psi_a}. \ee
\subsection{Infinitesimal limit}
To the end of this section we will informally derive infinitesimal
limit with some time independent vertex potential density. We will
do it by covering the space, let say $\mathbb{R}^d$, with lattice
and decrease its width to 0.

This time we would like that (informally):
$$P(\textrm{path}\ \gamma)\ \textrm{is proportional to}\ e^{-\int
V(\gamma(t))dt}$$ The problem is that such Brownian-like path has
infinite length - this probability distribution has to be thought
for example as a limit of made for some finite length
approximations of paths.

For simplicity we will take diagonal terms of $M$ equal 0, what
corresponds to moving with constant speed. But we will see later
that adding some constant on the diagonal doesn't change the
results.

This time $V$ is some (smooth) function
$\mathbb{R}^d\to\mathbb{R}$. Let's choose some time step
$\epsilon>0$ and cover the space with lattice of some width
$\delta>0$. We will treat it as a graph, in which each vertex is
connected with $2d$ neighbors.

We assume that the potential around a node is constant and equal
to the original potential $V$ in this node. One edge corresponds
to time $\epsilon$, so such discretized potential should be chosen
as $\epsilon$ times the original one.

This discretized model is symmetric, so $\phi=\psi$. \\The
eigenvalue relations ($M\psi=\lambda \psi$) for one-dimension
($d=1$) are:
$$\textrm{for all }i,\quad \psi_{i-1}e^{-\frac{\epsilon}{2}(V_{i-1}+V_i)}+
\psi_{i+1}e^{-\frac{\epsilon}{2}(V_i+V_{i+1})} =
\lambda_{\epsilon} \psi_i$$ where $\psi_i:=\psi(i\delta),\
2dV_i:=V(i\delta)$ (to simplify the final equation).

Now because $\epsilon\to 0$, we can use $e^{-\epsilon}\cong
1-\epsilon$:
$$\left(1-\frac{\epsilon}{2}(V_{i-1}+V_i)\right)\psi_{i-1}+
\left(1-\frac{\epsilon}{2}(V_i+V_{i+1})\right)\psi_{i+1}\cong
\lambda_{\epsilon} \psi_i$$
$$\frac{\psi_{i-1}-2\psi_i+\psi_{i+1}}{\epsilon}-
\frac{1}{2}\left((V_{i-1}+V_i)\psi_{i-1}+(V_i+V_{i+1})\psi_{i+1}\right)+\frac{2}{\epsilon}\psi_i\cong
\frac{\lambda_{\epsilon}}{\epsilon}\psi_i$$ It looks like there is
a problem with $\frac{2}{\epsilon}\psi_i$ term - it goes to
infinity. But it adds only a constant to the operator - it doesn't
change its eigenfunctions. The formula for the stochastic matrix
(\ref{sk}) also won't be affected - this time the powers will
become exponents and
$\frac{e^{t(\hat{M}+C)}}{e^{t(\lambda+C)}}=\frac{e^{t\hat{M}}}{e^{t\lambda}}$.
The $e^{t\lambda}$ term realizes the normalization of probability
distribution. So we will be able to ignore this constant term of
$\hat{M}$ later.

We have to connect $\delta$ and $\epsilon$. We see that to obtain
2nd derivative of $\psi$, we should take $\epsilon\sim \delta^2$,
what is characteristic for Brownian  motion. To simplify the final
equation, let's choose $\epsilon=2\delta^2$ time scale.

 Let's multiply above equation by -1 and assume some smoothness of $\psi$
and $V$ to write it for $\epsilon\to 0$:
$$-\frac{\psi_{i-1}-2\psi_i+\psi_{i+1}}{2\delta^2}+
2V_i\psi_i\cong \frac{2-\lambda_{\epsilon}}{\epsilon}\psi_i$$ We
can sum such equations for all dimensions and take the limit
$\epsilon\to 0$, getting the Schr\"{o}dinger equation:
$$\hat{H}\psi=-\frac{1}{2}\Delta\psi+V\psi=E_0\psi$$
where $\Delta:=\sum_i\partial_{ii}$,
$\hat{H}:=-\frac{1}{2}\Delta+V$.

 We've already explained that
we can ignore the constant term for $\hat{M}$, so we can take
$\hat{M}=-\hat{H}$.

 This time $E_0=\lim_{\epsilon\to
0}\frac{2d-\lambda_{\epsilon}}{\epsilon}$ is some new dominant
eigenvalue of differential operator $\hat{H}$, called Hamiltonian.
Now the largest $\lambda$ corresponds to the smallest $E$.
Corresponding eigenfunction $\psi$ should be obtained as the limit
of succeeding approximations - can be chosen as real nonnegative -
it's the ground state.

The equation for probability distribution becomes \be
p(x)=\psi^2(x)\qquad \textrm{where }\int_{\mathbb{R}^d}
\psi^2(x)dx=1\ee which looks similarly to known from the quantum
mechanics, but this time $\psi$ is real and nonnegative function.
Sometimes $\psi$ cannot be normalized, but it still can be used to
calculate the propagator.

The powers of $M$ matrix becomes the exponent of the differential
operator $\hat{M}$. The stochastic matrix becomes propagator,
which gives the probability density of finding a particle from $x$
after time $t$: \be
K(x,y,t)=\frac{<x|e^{-t\hat{H}}|y>}{e^{-tE_0}}\,\frac{\psi(y)}{\psi(x)}\ee

It's easy to check, that as we would expected:
$$\int_{\mathbb{R}^d} K(x,y,t)dy=1,\qquad \int_{\mathbb{R}^d} K(x,y,t)K(y,z,s)dy=K(x,z,t+s),$$
$$\int_{\mathbb{R}^d} p(x)K(x,y,t)dx = p(y).$$

As it was already mentioned, $e^{-tE_0}$ term is for
normalization. The $\psi$ division term corresponds to time
discretization, I think. Because $\psi$ should be continuous, for
small times the particle moves corresponding to the local
potential only. For larger times, this term starts to be important
- now the particle doesn't just move straightforward between these
points, but choose statistically some trajectory - this term
corresponds to the global structure of potential/topology of the
space.

\section{Asymmetric Numeral Systems (ANS)}
We will now show how to use found Markov process (or generally -
statistical algorithm) to deterministically encode some information.
Using the data, we have to generate succeeding symbols with given
probability distribution $(q_s)_{s=0,..,n-1}$.\\
To do it we could use any entropy coder, but in reversed order:
encoding into Markov's process correspond to decompression,
decoding to compression.\\

In practice there are used two approaches for entropy coding
nowadays: building binary tree (Huffman coding) and arithmetic
coding. The first one approximates probabilities of symbols with
powers of 2 - isn't precise. Arithmetic coding is precise. It
encodes symbol in choosing one of large ranges of length
proportional to assumed probability distribution ($q$).
Intuitively, by analogue to standard numeral systems - the symbol
is encoded on the most important position.
To define the actual range, we need to use two numbers (states).\\

We will construct precise encoding that uses only one state. It
will be obtained by distributing symbols uniformly instead of in
ranges - intuitively: place information on the least important
position. Standard numeral systems are optimal for encoding
streams of equiprobable digits. Asymmetric numeral systems is
natural generalization into other, freely chosen probability
distributions. If we choose uniform probability, with proper
initialization we get standard numeral system.

\subsection{General concept} We would like to encode an
uncorrelated sequence of symbols of known probability distribution
into as short as possible sequence of bits. For simplicity we will
assume that the the probability distribution is constant, but it
can be naturally generalized for various distributions. The
encoder will receive succeeding symbols and
transform them into succeeding bits.\\

 An symbol(event) of probability $p$ contains
$\lg(1/p)$ bits of information - it doesn't have to be a natural
number. If we just assign to each symbol a sequence of bits like
in Huffman coding, we approximate probabilities with powers of 2.
If we want to get closer to the optimal compression rates, we have
to be more precise. We see that to do it, the encoder have to be
more complicated - use not only the current symbol, but also
relate to the previous ones. The encoder has to have some state in
which is remembered unnatural number of bits of information. This
state in arithmetic coder are two numbers describing actual
range.\\

The state of presented encoder will be one natural number:
$x\in\mathbb{N}$.  For this subsection we will forget about
sending bits to output and focus on encoding symbols. So the state
$x$ in given moment is a natural number which encodes all already
processed symbols. We could just encode it as a binary number
after processing the whole sequence, but because of it's size it's
completely impractical. In 3.4 it will be shown that we can
transfer the youngest bits of $x$ to assure that it stays in the
fixed range during the whole process. For now we are looking for a
rule of changing the state while processing a symbol $s$:
\begin{equation}
 (s,x)\begin{array}{cccc}\textrm{encoding}\\\longrightarrow\\\longleftarrow\\\textrm{decoding}
 \end{array}x'
\end{equation}
So our encoder starts with for example $0$ and uses above rule on
succeeding symbols. These rules are bijective, so that we can
uniquely reverse whole process - decode the final state back into
initial sequence of symbols in reversed order.

In given moment in $x$ is stored some unnatural number of bits of
information. While writing it in binary system, we would round
this value up. To avoid such approximations, we will use
convention that $x$ is the
possibility of choosing one of $\{0,1,..,x-1\}$ numbers, so $x$ contains exactly $\lg(x)$ bits of information.\\

For assumed probability distribution of $n$ symbols, we will
somehow split the set $\{0,1,..,x-1\}$ into $n$ separate subsets -
of sizes $x_0,..,x_{n-1}\in\mathbb{N}$, such that
$\sum_{s=0}^{n-1} x_s=x$. We can treat the possibility of choosing
one of $x$ numbers as the possibility of choosing the number of
subset($s$) and then choosing one of $x_s$ numbers. So with
probability $q_s=\frac{x_s}{x}$ we would choose $s$-th subset. We
can enumerate elements of $s$-th subset from $0$ to $x_s-1$ in the
same order as in the original enumeration of $\{0,1,..,x-1\}$.

Summarizing: we've exchanged the possibility of choosing one of
$x$ numbers into the possibility of choosing a pair: a symbol($s$)
with known probability distribution ($q_s$) and the possibility of
choosing one of $x_s$ numbers. This ($x \rightleftharpoons
(s,x_s)$) will be the bijective coding we are looking for.

We will now describe how to split the range. In arithmetic coding
approach (Range Coding), we would divide $\{0,..,x-1\}$ into
ranges. In ANS we will distribute these subsets
uniformly.\\

We can describe this split using \textbf{distribution function}
$D_1:\mathbb{N}\to\{0,..,n-1\}$:
$$\{0,..,x-1\}=\bigcup_{s=0}^{n-1} \{y\in\{0,..,x-1\}:D_1(y)=s\}$$
We can now enumerate numbers in these subsets by counting how many
are there smaller of them in the same subset: \be \label{C2}
x_s:=\#\{y\in\{0,1,..,x-1\},\ D_1(y)=s\}\quad\quad\quad
D_2(x):=x_{D_1(x)}\ee getting bijective \textbf{decoding
function}(D) and it's inverse \textbf{coding function} (C):
$$D(x):=(D_1(x),D_2(x))=(s,x_s)\quad \quad C(s,x_s):=x.$$

Assume that our sequence consist of $n\in\mathbb{N}$ symbols with
given  probability distribution $(q_s)_{s=0,..,n-1}$ $\
\left(\forall_{s=0,..,n-1}\ q_s>0\right)$. We have to construct a
distribution function and coding/decoding function for this
distribution: such that \be\forall_{s,x}\quad\quad x_s \ \
\textrm{is approximately}\ \ x\cdot q_s.\ee We will now show
informally how essential above condition is. In 3.3 and 3.5 will
be shown two ways of making such construction.\\

Statistically in a symbol is encoded $H(q):=-\sum_s q_s \lg{q_s}$
bits.\\ ANS uses $\lg(x)-\lg(x_s)=\lg(x/x_s)$ bits of information
to encode a symbol $s$ from $x_s$ state. Using second Taylor's
expansion of logarithm (around $q_s$), we can estimate that our
encoder needs at average:
$$-\sum_s q_s \lg\left(\frac{x_s}{x}\right)\approx -\sum_s q_s
\left(\lg(q_s)+\frac{x_s/x-q_s}{q_s\ln(2)}-\frac{(x_s/x-q_s)^2}{2q_s^2\ln(2)}\right)=$$
$$=H(q)+\frac{1-1}{q_s\ln(2)}+\sum_s \frac{(x_s/x-q_s)^2}{2q_s\ln(2)}\ \ \ \ \textrm{bits/symbol.}$$
We could average \be\label{acc} \frac{1}{2\ln(2)}\sum_s
\frac{q_s}{x^2}(x_s/q_s-x)^2 =\frac{1}{\ln(4)}\sum_s
\frac{q_s}{x^2}\left(x_s/q_s-C(s,x_s)\right)^2 \ee over all
possible $x_s$ to estimate how many bits/symbols we are wasting.
\subsection{Asymmetric Binary System (ABS)}
 It occurs that in the binary case, we can find simple explicit
formula for coding/decoding functions.\\

We have now two symbols: $"0"$ and $"1"$. Denote $q:=q_1$, so $\tilde q:=1-q=q_0$.\\
To get $x_s\approx x\cdot q_s$, we can for example take \be
x_1:=\lceil xq
\rceil\quad\quad\quad\quad\quad\quad\left(\textrm{or
alternatively}\ x_1:=\lfloor xq \rfloor\right)  \ee \be
x_0=x-x_1=x-\lceil xq \rceil
\quad\quad\quad\quad\quad\left(\textrm{or}\ x_0=x-\lfloor xq
\rfloor \right)\ee Now using (\ref{C2}): $D_1(x)=1\
\Leftrightarrow$ there is a jump of $\lceil xq \rceil$ after it:
\be s:=\lceil (x+1)q \rceil-\lceil xq \rceil
\quad\quad\quad\left(\textrm{or}\ s:=\lfloor (x+1)q
\rfloor-\lfloor xq \rfloor\right) \ee
We've just defined \textbf{decoding} function: $D(x)=(s,x_s)$.\\

For example for $q=0.3$:\\

\begin{tabular}{|c||c|c|c|c|c|c|c|c|c|c|c|c|c|c|c|c|c|c|c|}\hline
$x$&0&1&2&3&4&5&6&7&8&9&10&11&12&13&14&15&16&17&18
\\\hline\hline $x_0$&&0&1&&2&3&&4&5&6&&7&8&&9&10&&11&12\\\hline
$x_1$&0&&&1&&&2&&&&3&&&4&&&5&&\\\hline
\end{tabular}\\
 \\

We will find coding function now: we have $s$ and $x_s$ and want to find $x$.\\
Denote $r:=\lceil xq \rceil-xq\in[0,1)$\\
$s:=\lceil (x+1)q \rceil-\lceil xq \rceil=\lceil (x+1)q -\lceil xq
\rceil\rceil=\lceil (x+1)q-r-xq\rceil=\lceil q-r \rceil$\\
$$s=1\Leftrightarrow r<q$$
\begin{itemize}
\item $s=1$: $\quad x_1=\lceil xq \rceil=xq+r$\\
$x=\frac{x_1-r}{q}=\Big\lfloor\frac{x_1}{q}\Big\rfloor\quad$
because it's natural number and $0\leq r<q$.
\item $s=0$: $q\leq r<1$ so $\tilde{q}\geq 1-r>0$\\
$x_0=x-\lceil xq \rceil=x-xq-r=x\tilde{q}-r$
$$x=\frac{x_0+r}{\tilde{q}}=\frac{x_0+1}{\tilde{q}}-\frac{1-r}{\tilde{q}}=
\Big\lceil\frac{x_0+1}{\tilde{q}}\Big\rceil-1$$
\end{itemize}
Finally \textbf{coding}: \be \label{decoding} C(s,x)=\left\{
\begin{array}{ll}
\Big\lceil\frac{x+1}{1-q}\Big\rceil-1 &\ \textrm{if}\ s=0\\
\ \Big\lfloor\frac{x}{q}\Big\rfloor &\ \textrm{if}\ s=1\end{array}
\right. \quad\quad\quad\left(\textrm{or}\ = \left\{
\begin{array}{ll}
\ \ \ \Big\lfloor\frac{x}{1-q}\Big\rfloor &\ \textrm{if}\ s=0\\
\Big\lceil\frac{x+1}{q}\Big\rceil-1 &\ \textrm{if}\ s=1\end{array}
\right.\right)\ee For $q=1/2$ it's usual binary system (with
switched digits). \subsection{Stream coding/decoding} Now we can
encode into large natural numbers ($x$). We would like to use
ABS/ANS to encode data stream - into potentially infinite sequence
of digits(bits) with expected uniform distribution. To do it we
can sometimes transfer a part of information from $x$ into a digit
from a standard numeral system to enforce $x$ to
stay in some fixed range ($I$).\\

Let the data stream be encoded as $\{0,..,b-1\}$ digits - in
standard numeral system of base $b\geq 2$. In practice we use
binary system ($b=2$), but thanks of this general approach, we can
for example use $b=2^8$ to transfer whole byte at once. Symbols
contain correspondingly $\lg(1/q_s)$ bits of information. When
they cumulate into $\lg b$ bits, we will transfer full digit
to/from
output, moving $x$ back to $I$.\\

Observe that taking interval in form ($l\in\mathbb{N}$): \be
I:=\{l,l+1,..,bl-1\}\ee for any $x\in\mathbb{N}$ we have exactly
one of three cases:
\begin{itemize}
  \item $x\in I$ or
  \item $x>bl-1$, then $\exists!_{k\in\mathbb{N}}\ \lfloor x/b^k\rfloor\in
  I$ or
  \item $x<l$, then $\forall_{(d_i)\in \{0,..,b-1\}^\mathbb{N}}\
  \exists!_{k\in\mathbb{N}}\ xb^k+d_1 b^{k-1}+..+d_k\in I$.
\end{itemize}
We will call such intervals \textbf{$b$-absorbing}: starting from
any natural number $x$, after eventual a few reductions
($x\to\lfloor x/b\rfloor$) or placing a few youngest digits in $x$
($x\to
xb+d_t$) we would finally get into $I$ in unique way.\\

For some interval($I$), define \be I_s=\{x:C(s,x)\in I\},\quad
\textrm{so that}\ I=\bigcup_s C(s,I_s).\ee Define \textbf{stream
decoding}:\\
\verb"{(s,x)=D(x);"\\
\verb" use s;       "(for example to generate symbol)\\
\verb" while(x"$\notin I$\verb")  x=xb+'digit from input' }"\\

\textbf{Stream coding}(\verb"s"):\\
\verb"{while(x"$\notin I_s$\verb")"\\
\verb"   {put mod(x,b) to output; x="$\lfloor$\verb"x/b"$\rfloor$\verb"}"\\
\verb" x=C(s,x) }"\\

\begin{figure}[h]
    \centering
        \includegraphics{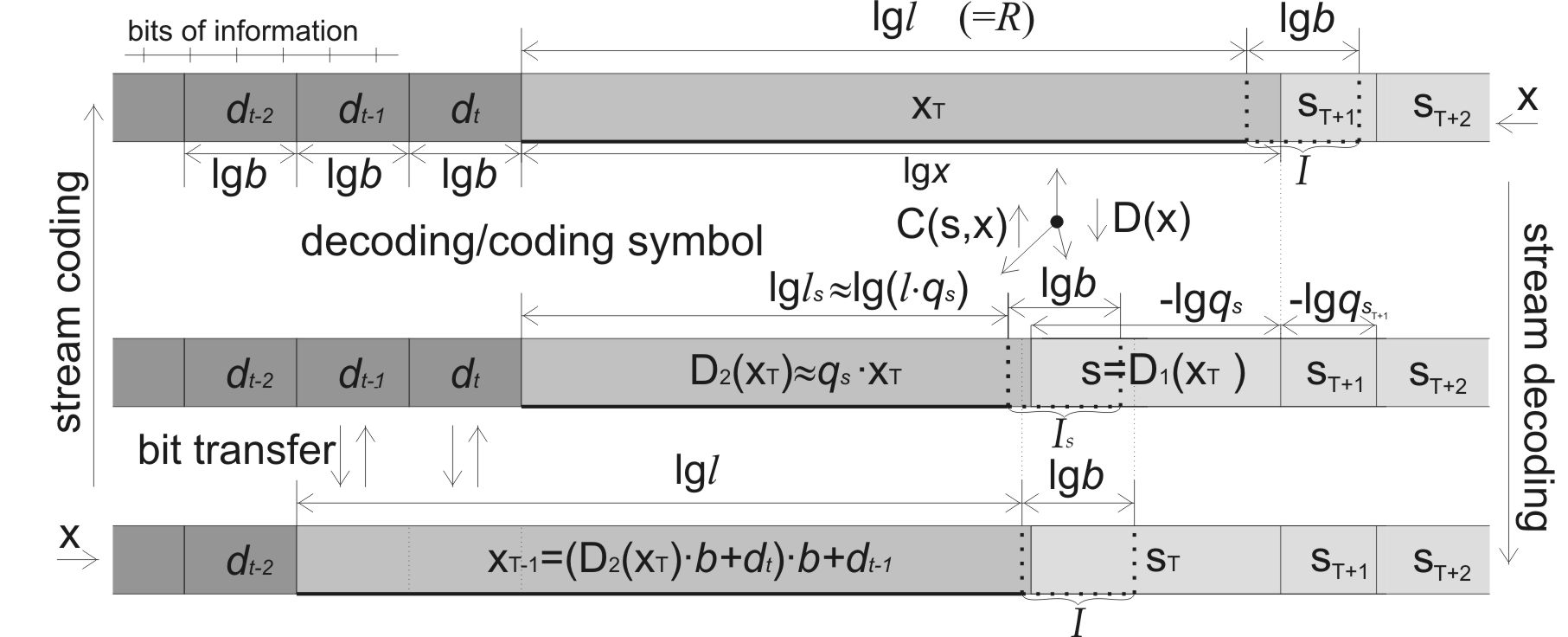}
        \caption{Stream coding/decoding}
\end{figure}

We need that above functions are ambiguous reverses of each other.\\
Observe that we would have it iff $I_s$ for $s=0,..,n-1$ and $I$
are $b$-absorbing: \be I=\{l,..,lb-1\}\quad\quad I_s=\{l_s,..,l_s
b-1\} \ee
 for some $l,l_s\in \mathbb{N}$.\\

We have: $\sum_s l_s(b-1)=\sum_s\#I_s=\# I=l(b-1)$.\\
Remembering that $C(s,x)\approx x/q_s$, we finally have: \be
l_s\approx lq_s\quad\quad \sum_s l_s=l.\ee

We will look at the behavior of $\lg x$ while stream coding $s$
now: \be \lg x\rightarrow\ \sim\ \lg x +\lg (1/q_s)\qquad
(\textrm{modulo}\ \lg b)\ee We have three possible sources of
random behavior of $x$:
\begin{itemize}
\item we choose one of symbol (behavior) in statistical(random) way,
\item usually $\frac{\lg q_s}{\lg b}$ are irrational,
\item $C(s,x)$ is near but not exactly $x/q_s$.
\end{itemize}

It suggests that $\lg x$ should cover uniformly the possible
space, what agrees with statistical simulations. That means that
the probability that we are visiting given state $x$ should be
approximately proportional to $1/x$.

Thanks of this observation we can for example estimate the number
of wasted bits/symbol (\ref{acc}). Mean value of
$(x/q_s-C(s,x))^2$ depends on used distribution function - for
precise functions like in ABS, it's smaller than 1. More important
is the mean value of $\frac{1}{x^2}$ - it should be about (by
integrating) $\frac{1}{l^2}\frac{b^2-1}{b^2}\ln{b}$. We can
manipulate $l$ and $b$ parameters to achieve wanted compromise
between speed and
precision of our encoder.\\

We will now focus on the \textbf{stream version of ABS}.\\
For practical reason we can take: \be l=2^R\quad\quad
b=2^w\quad\quad I=\{2^R,..,2^{R+w}-1\}\ee We have to check when
$I_0, I_1$ are $2^w$-absorbing.\\ Observe that $D_2(bl)\in
\{bl-\lceil blq \rceil, \lceil blq \rceil\}$ have to be the
smallest number above correspondingly $I_0$ or $I_1$ - have to be
equal
$bl_0$ or $bl_1$.\\
In both cases $I_0, I_1$ are $2^w$-absorbing iff \be \label{cond}
\lceil blq \rceil=\lceil 2^{R+w}q \rceil \quad \textrm{is a
multiplicity of}\ 2^w\ee So if we want to use formulas explicitly,
the precision of $q$ should be at most $R$
bits.\\

In implementation of data compressor using ABS, we can:
\begin{itemize}
  \item calculate formulas for every symbol while processing data -
  it is more precise and we can transfer a few bits at once,
but it can be a bit slower, and we need to be careful with
(\ref{cond}), or
\item store the coding tables in memory - smaller precision, needs memory and time for
initialization, but should be faster and we have large freedom in
choosing coding/decoding functions. For example by changing a few
last symbols we can repair (\ref{cond}) for more precise $q$.
\end{itemize}

Practical problem is that decoded and encoded sequences of symbols
are in reverse order - to use probability prediction methods, we
have to make predictions to the end, than encode in backward
order.
Now decompression is straightforward.\\
In Matt Mahoney's implementations (fpaqb, fpaqc on \cite{mah}) the
data is divided into compressed separately segments, for which we
store $q$ from the prediction process.

\subsection{Asymmetric Numeral Systems(ANS)} In the general case:
encoding a sequence of symbols with probability distribution
$q_0,..,q_{n-1}$ for some $n>2$, we could divide the selection of
symbol into a few binary choices and just use ABS. We will see
that we can also encode such symbols straightforward.
Unfortunately I couldn't find practical explicit formulas for
$n>2$, but we can calculate coding/decoding functions while the
initialization, making processing of the data stream extremely
fast. The problem is that we rather cannot table all possible
probability distributions - we have to initialize for a few and
eventually
reinitialize sometimes.\\

Fix some $l,b,(q_s)$ and choose some $l_s\in\mathbb{N}:\
l_s\approx
lq_s,\ \sum_s l_s=l$.\\
We have to choose the distribution function ($D_1$) for $x$ from
$l$ to $bl-1$ - distribute $(b-1)l_0$ appearances of symbol
'0',$(b-1)l_1$ of '1', ... , $(b-1)l_{n-1}$ of '$n-1$'.

We could do it using that $D_2(x)$ is about $xq_s$ as previously
to choose the most appropriate symbol for succeeding $x$. It would
require priority queue for symbols.

In this section we will focus on a bit less precise but faster
statistical method: fill a table of size $(b-1)l$ with proper
number of appearances of symbols and for succeeding $x$ take
symbol of
random number from this table, reducing the table.\\
Another advantage of this approach is that after fixing $(l_s)$,
we still have huge (exponential in $\#I$) number of possible
coding functions - we can choose one using some key, additionally
encrypting the data.\\

\textbf{Initialization}: choose some $l_s\in\mathbb{N}:\ l_s\approx lq_s,\ \sum_s l_s=l$;\\
\verb"m=(b-1)l; symbols"
=$(\overbrace{0,0,..,0}^{(b-1)l_0},\overbrace{1,1,..,1}^{(b-1)l_1},..,
\overbrace{n-1,..,n-1}^{(b-1)l_{n-1}})$;\\
\verb"For x=l to b*l-1"\\
\verb"  {i=random natural number from 1 to m;"\\
\verb"   s=symbols[i]; symbols[i]=symbols[m]; m--;"\\
\verb"   D[x]=(s,"$l_s$\verb") " or \verb" C[s,"$l_s$\verb"]=x"\\
\verb"   "$l_s$\verb"++}"\\

Where we can use practically any deterministic pseudorandom number
generator, like Mersenne Twister(\cite{mer}) and use eventual key for its initialization.\\
Practically without any cost we can increase the preciseness of
this
algorithm as much as we want by dividing $I$ into subsegments initialized separately.\\

Modern random number generators are practically unpredictable, so
the ANS initialization would be. It chooses for each $x\in I$
different random local behavior, making the state practically
unpredictable hidden variable.

Encryption based on ANS instead of making calculation while taking
succeeding blocks as standard ciphers, makes all calculations
while initialization - processing of the data is much faster: just
using the tables.

 Another advantage of preinitialized cryptosystem
is that it's more resistant to brute force attacks - while taking
a new key to try we cannot just start decoding as usual, but we
have to make whole initialization earlier, what can take as much
time as the user
wanted.\\

We can use this approach as an error correction method also. For
example by introducing a new symbol - the forbidden one and
rescaling the probability of the allowed ones. Now if this symbol
occurs, we know that there was an error nearby. Knowing the
statistical error distribution model, we can create a long list of
possible errors, sorted by the probability. Now we can try to
correct as it would be this case for succeeding points of this
list and verify by trying to encode the following message. In this
way we use that the most of blocks are correct, so we can transfer
surpluses of unused redundancy to help with pessimistic cases.

We can also use huge freedom of choice for ANS to make the
correction easier and faster - for example by enforcing that two
allowed symbols has Hamming distance at least some number. We can
for example get Hamming code this way as degenerated case - each
allowed symbol appears once, so the intermediate state is just 1 -
we don't connect redundancy of blocks. In other cases we can use
the connection between their redundancy to cope with badly damaged
fragments.

\section{Information capacity of general models}
In this section we will give formalism and methodology for general
models and prove existence of entropy for for large class of them.
\subsection{Basic definitions}
We will operate on $f:A\to \mathcal{A}\ $ ($A\subset X$)
functions. They will be called as before \emph{patterns} and their
domain, denoted $\mathcal{D}(f)$ - their \emph{shape}.
\begin{df} Model \emph{will be called a triple}
$(X,\mathcal{A},\mathcal{C})$\\
space $X\subset\mathbb{Z}^m$\\
alphabet $\mathcal{A} $\emph{  -finite set of symbols  } \\
constrains \emph{(by forbidden valuations):}
$\mathcal{C}\subset\{c:A\to \mathcal{A}:\ A\subset X\}$\\
\emph{then} elements of model \emph{are }$V(X)$, \emph{where for}
$A\subset X$ \emph{its} valuations \emph{are:}
$$ \qquad V(A):=\{v:A\to\mathcal{A}\vert\forall_{c\,\in \mathcal{C}:\, \mathcal{D}(c)\subset A}
\ \exists_{x\in \mathcal{D}(c)} c(x)\neq v(x)\}\quad\quad
V:=\bigcup_{A\in X:\ \#A<\infty} V(A)$$
\end{df}

Digression: it is a very general definition. Instead of defining
by forbidden states, we could do it by allowed ones (for example -
take large enough subset of $X$ and take all allowed valuations).

We can write for example tiling problems in that formalism (denote
each tile with a separate symbol and put their shapes in "allowed
constrains"). We know that there can happen very different
situations - even enforcing nonperiodic tiling.

To control our model we will have to limit this class.\\

The main example we will use is the Hard Square model.
\begin{df}Hard-square model (HS):
\begin{displaymath}
 (X=\mathbb{Z}^2,\mathcal{A}=\{0,1\},\mathcal{C}=\{(((i,j),1),((k,l),1)):i,j,k,l\in\mathbb{Z},|i-k|+|j-l|=1\})
\end{displaymath}
\end{df}

In each node of two-dimensional lattice we put $0$ or $1$ such
that there are no neighboring '1'. It is one of the simplest
models which haven't been solved analytically. In \cite{bax} we
can find exact formula for entropy of similar Hard Hexagon Model
(we add upper left and lower right to the neighborhood), but this
methodology creates some unsolvable singularities for HS. In
\cite{bax1} 43 digits of entropy of HS are found - I will use it
to evaluate algorithms.

Unfortunately this methodology cannot be
used to find statistics.\\

Define:
\begin{df}
\ \\
Neighborhood of $x\in X$: $\quad N_x:=\{y\in X:\exists_{c\in
\mathcal{C}}\
x,y\in \mathcal{D}(c)\}$\\
Range of constrains: $\quad L:=\max\{|y_i-x_i|:x\in X,\ y\in N_x,\ i\in\mathbb{N}\}$\\
Interior of  $A\subset X$ : $\quad A^-:=\{x\in A:N_x\subset A\}$\\
Boundary of $\quad A\subset X$ : $A^o:=A\backslash A^- $ \\
Thickening of $A\subset X$:$\quad  A^+:=\bigcup_{x\in A} N_x$\\
\emph{We will call set $A\subset X$ }connected, \emph{if}
$$\forall_{x,y\in A}\exists_{n\in\mathbb{N},z^0,...,z^n\in X}\quad z^0=x,\ z^n=y,\
\forall_i \sum_j |z^i_j-z^{i+1}_j|=1.$$
\end{df}
Of course $A^{-+}\subset A\subset A^{+-}$.\\

For translative invariant models, we need to know only the
neighborhood around one point, e.g. for HS:$\ N_x=x+\{(0,0),(1,0),(0,1),(-1,0),(0,-1)\}$.\\

Digression: neighborhood is always symmetric set ($x\in
N_y\Leftrightarrow y\in N_x$), $x\in N_x$, so
$$\rho
(x,y)=\min_{n}\exists_{(x^0=x,x^1,...,x^n=y)}\ \forall_i
x^{i+1}\in N_{x_i}$$ is a natural metric for a given model.\\

HS model has many symmetries: generated by translations, axial
symmetry and rotations by $\pi/2$.
\begin{df} \emph{A bijection }$S:X\hookrightarrow X$ \emph{will be called }symmetry \emph{if}\\
\begin{equation} c\in \mathcal{C}\Leftrightarrow c\circ S\in \mathcal{C}
\end{equation}
\end{df}
\subsection{Existence of entropy}
We will now reduce the family of models we will focus on and prove
the existence of positive average entropy for them.\\

 \textbf{For the rest of this paper we assume, that:}
\begin{itemize}
\item $X=\mathbb{Z}^m$
\item has finite range of constrains
\item is translational invariant - every translation is its symmetry($t_x(y):=y+x$)
\item is simple($\#$):
\end{itemize}

\begin{df} \emph{We call model }simple($\#$),
\emph{if $\#V(X)>1$ and \\ there exists natural numbers $N>n\geq
0$, such that}
$$\forall_{\mathrm{connected }\ A\subset X}\ \forall_{v\in V(\left(A^{N})^o\right)}
\ \forall_{u\in V(A^{n})}\ \exists_{w\in V(\left(A^N\right)^-\bs
A^n)}\quad u\cup w\cup v\in V(A^N)$$
\end{df}
where $A^0:=A,\ A^{i+1}:=(A^i)^+$.\\

In other words: for sets "nice" enough (of the form $A^n$), exists
thickening ($N-n$ times) that for any valuation of its boundary,
every valuation of $A^n$ is still allowed.

For example models with neutral symbol - which isn't in any
constrain (we can always use it: $0$ in HS) are simple(\#):
$n=0,\ N=2\ $ - we fill $A^+\bs A $ with this symbol.\\

To the end of this paper we use notation: \be N^A_u\equiv
N(A,u):=\#\{w\in V(A):w|_{\mathcal{D}(u)}=u\},\quad N(A)\equiv
N\left(A,\{\}\right).\ee
\begin{lem}\label{lem1}
Denote \emph{$k$-block}: $\beta_k:=\{0,1,...,k-1\}^m$, \emph{now:}
$$\exists_{n'\geq \max(L,1)}\forall_{v\in V(\beta_{n'}^+\bs \beta_{n'})}
N(\beta_{n'}^+,v)>1.$$
\end{lem}

Simply speaking: for every valuation of the neighborhood of large
enough block we can valuate it in more than one way.\\
\emph{Proof:} Because $\#V(X)>1$, we can choose $k$: $\#V(\beta^k)>1$.\\
 $\beta_k^{N-1}$ can be placed in some $n'$ - block ($n'\geq L$).     $\qquad\qquad\qquad\Box$

\begin{tw}\label{tw1}
For models as above, there exists entropy ($H$) and is positive:\\
There exists increasing sequence of sets $(A_i)$: $A_i\subset X,\
\#A_i<\infty,\ \bigcup_i A_i=X$, such that there exists a limit:
$$0<H:=\lim_{i\to\infty}\frac{lg(N(A_i))}{\#A_i}\leq \lg \#\mathcal{A}$$
\end{tw}
\emph{Prove:}\\
Take $n'$ like in the lemma. We will operate on \emph{blocks}:
$B\equiv \beta_{n'}$ placed in nodes of lattice $n'\mathbb{Z}^m$.\\
Because $n'\geq L$, valuations of a block can be restricted only by
valuations of adjoined blocks: $B+n'(\{-1,0,1\}^m\bs\{0\})$.\\

\begin{figure}[h]
    \centering
        \label{szk}
        \includegraphics[width=\textwidth]{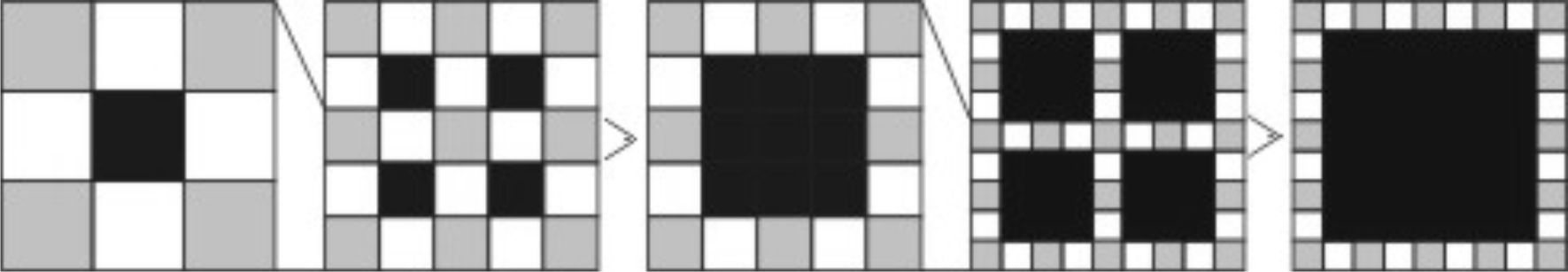}
        \caption{Block division for $A_1,\ C'_2,\ A_2,\ C'_3,\ A_3$. Black - "essence".
        The rest (filler) we valuate such that all blocks denoted with the same color has the same pattern.}
\end{figure}

First of all we will find the filling pattern - periodic valuation
of the space.

Take the lattice $Y:=2n'\mathbb{Z}^m$ and numerate anyhow
$\{0,1\}^m=\{x^i\}_{i=1,...,2^m}.$\\
Now every block from $Y+B+x^1$ can be valuated independently from
the other ($n'\geq$ range of constrains). Choose any $w_1\in V(B)$.\\
Now using lemma \ref{lem1}, after valuating these blocks we can
find some $w_2\in V(B)$ for $Y+B+x^2$ not colliding with all
$w_1$. And so on we get the universal periodic filling: $$ V(X)\ni
w:\forall_{y\in Y, x\in B, i\in\{1,..,2^m\}}\quad
w(y+x+x^i):=w_i(x).
$$

Now we can go to the main construction.\\
As the sequence we are looking for, let's take $$A_i:=B+n'
\{0,...,2^i\}^m,\quad\ H_i:=\frac{\lg(N(A_i))}{\#A_i}$$
Notice that $H_i\leq\#\mathcal{A}$ - we would get this entropy without constrains.\\
To prove that $H_i$ has a limit, we will construct increasing
sequence $H'_i$, that $\forall_i\ H'_i\leq H_i\leq\#\mathcal{A}$ and
show that $\lim_{i\to\infty}H'_i-H_i=0$. \\
Monotone and bounded sequence has a limit, so $H_i$ will have the same.\\

We will make some construction to ensure the monotonicity:
$$C_i:=B+n'\{x\in\{0,...,2^i\}^m:\exists_i
x_i\in\{0,2^i\}\}$$
external blocks of $A_i$ which will be filled with $w$.\\
We will need intermediate step:
$$C'_{i+1}:=B+n'\{x\in\{0,...,2^i\}^m:\exists_i
x_i\in\{0,2^i,2^{i+1}\}\}$$ \be H'_i=
\frac{\lg(N(A_i,w|_{C_i}))}{\#A_i}\quad\quad H''_{i+1}=
\frac{\lg(N(A_{i+1},w|_{C'_{i+1}}))}{\#A_{i+1}} \label{eentr}\ee
Of course $H''_i\leq H'_i\leq H_i$.\\
Now the "essence" of $A_{i+1}\bs C'_{i+1}$ is made exactly of $2^m$
"essences" from the previous step,
with the same valuation of surrounding blocks.\\
So $N(A_{i+1},w|_{C'_{i+1}})=2^m N(A_i,w|_{C_i})$,
$\#A_{i+1}=\left(\frac{2^{i+1}+1}{2^i+1}\right)^m \#A_i<2^m\#A_i$\\
We get: $H'_i<H''_{i+1}$,
$$H'_i<H''_{i+1}\leq H'_{i+1}<...\leq \#\mathcal{A}.$$

There left only to show, that $\lim_{i\to\infty} H_i-H'_i=0$\\
Look at $D_i:=(A_i\bs
C_i)^{\overbrace{--...-}^{N-1}\overbrace{++...+}^n}$.\\
($A^{-+}\subset A$), so from ($\#$): \\
we can freely valuate $D_i$, independently to valuation on $C_i$.\\
Because $\beta_k^-\supset \beta_{k-2L}+(l,l,...,l)$, so
$\beta_{2^i-2NL}+(NL,...,NL)\subset D_i$, \\for large $i$:
$$\# D_i\ \sim (2^i)^m\qquad\#(A_i\bs D_i) \sim (2^i)^{m-1}$$
So $\lim_{i\to\infty}\frac{\#(A_i\bs D_i)}{\#D_i}=0$.\\

We have $N(D_i)\leq N(A_i,w|_{C_i}\leq N(A_i)$\\
But $\frac{N(A_i)}{N(D_i)}\leq (\#\mathcal{A})^{\#(A_i\bs
D_i)}$ - equality would be without constrains.\\
So $\frac{\lg N(A_i)-\lg N(D_i)}{\#A_i}\leq\frac{\#(A_i\bs
D_i)\lg(\#\mathcal{A})}{\#A_i}\to 0$.\\

From three sequences: $\lim_{i\to\infty} H_i-H'_i=0\quad\quad\quad\quad \Box$.\\
For large class of models we can speak about their entropy.\\

For the rest of this work we add assumption of
\emph{irreductibility}:
\begin{df}
\emph{We call translational invariant model }irreducible, \emph{if:}
$$\bigcup_i\  \{0\}^i=X$$
\end{df}
Where $A^0:=A,\quad A^{i+1}:=(A^i)^+$.\\

If a model isn't irreducible (is reducible): $Y:=\bigcup_i
\{0\}^i\neq X$\\
Because $Y=-Y,\ Y=Y+Y$, so $y\in Y\Rightarrow y\mathbb{Z}\subset
Y$ - $Y$ is a periodic lattice - there exists linearly independent
$y^1,..,y^{m'}\in X$, such that:
$$Y=\sum_{i=1,...,m'} \mathbb{Z}y^i$$
Now if we make a transformation: $y^i\to
(\delta_{ij})_{j=1,...,m'}$ we get corresponding $m'$-dimensional
irreducible model.

$x\sim y\Leftrightarrow x-y\in Y$ is equivalence relation, so
$X=\bigcup_i x^i+Y$ (disjoint sum for some $(x_i)$) can be treated
as identical, independent lattices.

 The average entropy is the same as for the reduced model.

\section{Statistical approach}
Now we will want to find \emph{optimal statistical description},
by averaging over all valuations, like in one-dimensional case.
\subsection{Statistical description}
\begin{df} (Statistical) description\emph{ is a function:}
\begin{displaymath}
p:\{(A,f):A\subset X, \ \#A<\infty,\ f:A\to\mathcal{A}\}\to [0,1]
\end{displaymath}
such that
\begin{equation}
\forall_{A\subset X,\ \#A<\infty} \forall_{x\in X\setminus
A}\forall_{ f:A\to\mathcal{A}}\
\sum_{a\in\mathcal{A}}p_{f\cup\{(x,a)\}}=p_f \label{nor}
\end{equation}
\begin{displaymath}
p_{\{\}}=1
\end{displaymath}
\end{df}
where $p_f\equiv p(\mathcal{D}(f),f)$.\\

It gives for each shape $A$, the probability distribution of
valuations on this shape. Because of translational invariance, it
will be shown that $f$ doesn't depend on its position - for
example $p(01)$ denotes the probability that when choosing a node,
it and its right neighbor are valued correspondingly to 01.\\
The conditions (\ref{nor}) ensure normalization to 1. (e.g. p(01)+p(00)=p(0)).\\

Now we would like to take an average over elements, but we can
count only valuations on finite sets - we have to choose some
sequence of finite sets tending to the whole space.
\begin{df}\ \\
\emph{We call series of finite sets} normal sequence of sets
$(A_i)_{i\in \mathbb{N}}$ \emph{ if}
\begin{displaymath}
A_0\subset A_1\subset A_2\subset...
\end{displaymath}
\begin{displaymath}
\bigcup_{i\in\mathbb{N}}A_i=X.
\end{displaymath}

\emph{For $f,v\in V,\ A\subset X\ :\ \mathcal{D}(v)\cap
\mathcal{D}(f)=\emptyset,\
\mathcal{D}(f)\subset A,\ \mathcal{D}(v)\subset A$}\\
 \emph{we will call its }$(A,v)$ approximation of optimal description:
\be p^{A,v}_f:=\frac{N(A,v\cup f)}{N(A,v)}. \ee
\end{df}
Where $N(A,u)=\#\{w\in V(A):w|_{\mathcal{D}(u)}=u\}$. Denote: $p^A_f:=p^{A,\emptyset}_f$.\\

Now for some $B\supset A,\ v\in V(B^o),\ D_f\subset A^-,\ f\in
V(A)$:
\begin{eqnarray}
p^{B,v}_f&=&\frac{N(B,v\cup f)}{N(B,v)} =\frac{\sum_{u\in
V(A^o\backslash D_v)}N(B\backslash A^-,u\cup v)N(A,u\cup f)}{N(B,v)}=\nonumber\\
&=&\sum_{u\in V(A^o\backslash D_v)}p^{A,u}_f\frac{N(B\backslash
A^-,u\cup v)N(A,u)}{N(B,v)}=\sum_{u\in V(A^o\backslash
D_v)}p^{A,u}_f p^{B,v}_u\label{eq2}
\end{eqnarray}

Where we've divided the sum over all valuations into the sum
within and outside topologically separating set $A^o$.\\

 We want to find the optimal description $p_f^{\,o}$ as the limit of succeeding
approximations $p_f^A$ for sets from a normal sequence.  We've
just shown that we get approximation for the succeeding set, as
weighted average $\left(\sum_{u\in V(A^o\backslash
D_v)}p^{B,v}_u=1\right)$ of approximations from the previous one
for the same pattern $f$.

So going to the next set won't give worse approximation:
\begin{displaymath}
A\subset B\Rightarrow
\hat{p}^A_f\geq\hat{p}^B_f\geq\check{p}^B_f\geq\check{p}^A_f
\end{displaymath}
where
\begin{displaymath}
\check{p}^A_f:=\min_{v\in V(A^o)}\ p^{A,v}_f\qquad
\hat{p}^A_f:=\max_{v\in V(A^o)}\ p^{A,v}_f\qquad
d^A_f:=\hat{p}^A_f-\check{p}^A_f
\end{displaymath}

We've explained that $d^A_f$ isn't growing in normal sequence. If
we would prove that for some normal sequence $(A_i),\ d^{A_i}_f$
is decreasing to 0, than taking any other normal sequence $(B_i)$,
because $\forall_i\exists_j\ A_i\subset B_j$,
$\lim_{i\to\infty}p^{A_i}_f=\lim_{i\to\infty}p^{B_i}_f$.

So this limit would be the only reasonable optimal description.

Unfortunately I cannot prove formally its existence. I have to
assume it:
\begin{zal}[*]\label{zal}
For any pattern $f$, $d^{A_i}_f\to 0$ for some normal sequence
$(A_i)$.
\end{zal}

We can now define:
\begin{df}Optimal description, \emph{is } $p^{\,o}_f:=\lim_{i\to\infty}
p^{A_i}_f$\\ \emph{where} $(A_i)$ \emph{- any normal sequence.}
\end{df}
\subsection{Optimal description fulfills general Markov's property}
Standard Markov's property can be thought that knowing only both
ending symbols of a sequence, we know the probability distribution
of its interior. In the second section we'd shown that
one-dimensional optimal description fulfills Markov's property.
Now we will see that in the general case analogous property is
fulfilled - knowing symbols on the boundary of some set, we know
the probability distribution in its interior. As previously, it
will be uniform distribution among possible valuations.

Let's show before, that optimal description preserves symmetries\\
$S$ - a symmetry of model\\
$S':V(X)\to V(S(X))=V(X), S'(v)(x)=v(S^{-1}(x))$ - bijection on
$V(X)$ \\
Now for a normal sequence $(A_i)$ and some pattern $f$, take
(using freedom of choice) normal sequence $(B_i):B_i=S(A_i)$ and
pattern $S'(f)$: \beas p^o_f&=&\lim_{i\to\infty}\frac{\#\{w\in
V(A_i):w|_{\mathcal{D}(f)}=f\}}{\#\{w\in V(A_i)\}}=\\&=&
 \lim_{i\to\infty}\frac{\#\{S'(w)\in
 V(B_i):S'(w)|_{\mathcal{D}(S'(f))=S(\mathcal{D}(f))}=S'(f)\}}{\#\{S'(w)\in
 V(B_i)\}}=p^o_{S'(f)}
 \eeas
\begin{sps}
For any symmetry of model  $S$ and patter $f$ $$p^o_f=p^o_{S'(f)}.$$
\end{sps}

We will now discuss \emph{local optimality condition} (LOC).\\
While having finite space, we have finite ($N$) number of elements
- we can store there the largest number of information ($\lg
(N)$), if they have uniform probabilistic distribution - we don't
favor any. We have analog to this condition in infinite space:
when we valuate boundary of some finite set, all available
valuations of its interior are equally probable - its LOC.
\begin{sps} Optimality of statistical description $p$ is equivalent
to LOC(Local Optimality Condition):\par for any $B\subset X,\
\#B<\infty,\ A\subset B^-,\ f\in V(B)$:
$$p(f)=\frac{p(f|_{B\backslash A})}{N(B,f|_{B\bs A})}$$
\end{sps}
\emph{Proof:}
\begin{enumerate}
\item Take $p=p^o,\ (A_i)$ - normal sequence, $A,\ B,\ f$ like above
$$N(A_i,f|_{B\bs A})=\#\{u\in V(A):u\cup f|_{B\bs A}\in V(B)\}N(A_i,f)=N(B,f|_{B\bs
A})N(A_i,f)$$
$$
p^o_f=\lim_{i\to\infty}\frac{N(A_i,f)}{N(A_i)}=\lim_{i\to\infty}\frac{N(A_i,f|_{B\bs
A})}{N(B,f|_{B\bs A})N(A_i)}=\frac{p(f|_{B\backslash
A})}{N(B,f|_{B\bs A})}
 $$
\item assume now LOC for  $p$ : $B\subset X:\ \#B<\infty,\ \mathcal{D}(f)\subset A=B^-$
\be p(f)=\sum_{v\in V(B^o)}p(f\cup v)=\sum_{v\in V(B^o)}N(B,v\cup
f)\frac{p(v)}{N(B,v)}= \sum_{v\in V(B^o)}p(v) p^{B,v}_f
\label{eq1}\ee we get the first equality using normalization
(\ref{nor}), the second from LOC (we can fill $A\bs \mathcal{D}(f)$ in $N(B,v\cup f)$ ways)\\
Taking as $B$ succeeding elements of some normal sequence, we get
thesis. $\quad\Box$
\end{enumerate}

\begin{center}
  \textbf{Remarks}
\end{center}
\begin{enumerate}
\item For a pattern $f,\ A\subset X:\mathcal{D}(f)\subset A^-$ we have(\ref{eq1}):
$$p^o_f=\sum_{v\in V(A^o)}p^o_v p^{A,v}_f\label{wz1}$$

Now take for example sequence $A_0:={0},\ A_{i+1}:=A_i^+$, from
(*): $p^{A_i,v}_f\to p^o_f$ - dependence of probability
distribution of
distant nodes decrease:\\
\textbf{Assumption (*) is equivalent vanishing of long distance
correlations.}
\item LOC is equivalent \emph{pLOC - point local optimality condition}:
$$\forall_{B:N_0^o\subset B\subset X\bs \{0\},\ \#B<\infty}\forall_{v\in V(B)}
\forall_{a,b\in \mathcal{A}}\ p_{v\cup
\{(0,a)\}}=p_{v\cup\{(0,b)\}}$$ where $N_0^o=N_0\bs \{0\},\ N_0$ -
neighborhood of $0$.

It gives smaller set of conditions for optimality - we have only
to check LOC for $\#A=1$. We will use it later to generate
approximations of the uniform distribution over elements. It says
for example for HS that for any finite pattern:
$$p\left(\begin{array}{ccc}x&0&x\\0&0&0\\x&0&x\end{array}\right)=
p\left(\begin{array}{ccc}x&0&x\\0&1&0\\x&0&x\end{array}\right)$$
Where "$x$" denotes some valuations outside $N_0$.\\
formally:
\begin{large}
$$\ \forall_{A\subset X\bs N_0:\
\#A<\infty}\forall_{v\in V(A)}\ p_{f^*_0\cup v}=p_{f^*_1\cup v}$$
\end{large}
where $f^*:=0|_{N_0\bs \{0\}}\ ,\ f^*_a:=f^*\cup \{(0,a)\}$ \\
Other valuations of $N^o_0$ enforce 0 in the middle.\\

\emph{Proof:} by induction over $\#A$: for $\#A=1$ - pLOC\\
assume we've proved LOC for $\#A=k-1$\\
Take some $A,B,v:\#A=k,\ B\supset A^+,\ v\in V(B\bs A)$\\
We want to show that for any $f,g\in V(A)$
$$p_{v\cup f}=p_{v\cup g}$$
If there exists $x\in A$ such, that $f(x)=g(x)$, than we move $x$
to $B$ and use induction assumption. If not, we make intermediate
step to $f'=f\bs\{(x,f(x)\}\cup\{(x,g(x)\}$ for some $x\in
A$.$\quad\quad\Box$
\item Take a sequence  $A_0:=\{0\},\ A_{i+1}:=A_i^+$\\
than: $$\lim_{i\to\infty}\frac{\lg(N(A_i))}{-\sum_{v\in
V(A_i)}p^o_v\lg p^o_v}=1$$

\emph{Proof:} set $i\in\mathbb{Z}$\\
Take any valuation of $A_{i+1}^o$ , we have uniform distribution of
available valuations on $A_i$. Using ($\#$) ($n$ i $N$), we can
freely valuate $A_{i-N+n}$, so we have more valuations than
$N(A_{i-N+n})$:
$$-\sum_{v\in V(A_i)}p^o_v \lg p^o_v\geq \lg N(A_{i-N+n})$$
Now we repeat discussion from the end of proof of theorem \ref{tw1}
($\lim_{i\to\infty}\frac{\#A_{i-N+n}}{\#A_i}=1$) and get the
thesis.$\quad\quad\Box$
\end{enumerate}

So we've justified that \textbf{optimal description has the same
average capacity as the model.}
\subsection{Sequential statistical desription}
In statistical description we have some excessive information
because of the normality conditions. We can get rid of them in
e.g. two ways:
\begin{enumerate}
\item We can limit to patterns without fixed symbol ($a$). E.g. for HS, $s:A\to p_{0|_A}$\\
\emph{Proof:} We want to get probability of some $f$ with $k+1$
appearances of $a$, e.g.: $f(x)=a$, now:
$$p_f=p_{f\bs \{(x,a)\}}-\sum_{b\in\mathcal{A}\bs\{a\}} p_{f\bs
\{(x,a)\}\cup\{(x,b)\}}\ .$$
\item \emph{Sequential description :}\\
Take any numeration of nodes of the space:
$\{x^i\}_{i=0,1,...\infty}=X$\\
and fix some $a\in\mathcal{A}$
$$\forall_{v=(v_0,v_1,...,v_{k-1})\in\mathcal{A}^k}\
\forall_{b\in\mathcal{A}\bs\{a\}} \quad q_b(v):= \frac{p_{f_v\cup
\{(x_k,b)\}}}{p_{f_v}}$$
 where
$\mathcal{D}(f_v):=\{x^i\}_{i=0,...,k-1},\ f_v(x^i):=v_i$.

We are taking successively $x^i$ and using valuation of previous
nodes we get its probability distribution of valuations .\\

\emph{Proof:} We want to find some $p_f$,\\ $\exists_k\
\{x_0,...,x_k\}\supset \mathcal{D}(f),\ A=\{x_0,...,x_k\}$
$$\forall_{u\in V(A)}\  p_u=\prod_{i=0}^k q_{u(x_i)}((u(x_0),...,u(x_{i-1}))
\qquad\left(q_a(w)=1-\sum_{b\in\mathcal{A}\bs\{a\}}\ q_b(w)\right)$$
$$p_f=\sum_{u\in V(A\bs \mathcal{D}(f)):u\cup f\in V(A)} p_{f\cup u}\ .$$

In the next section we will forget about assumption that there are
only finite number of previous nodes.
\end{enumerate}
So we are able to generate elements with given statistical
description - visit successively $x^i$ and generate its valuation
with appropriate distribution. \\

Digression: assume now that we have some element
$f:X\to\mathcal{A}$ generated from uniform distribution (e.g.
using optimal algorithm). For a given shape $A$, the value of
$f|_A$ should statistically correspond to optimal statistical
description. So assuming vanishing of long-range correlation, we
should get optimal description just by "averaging" $f|_{A+x}$ over
$x\in nX$ where $n$ is some large number. The same
would be for any translations of $nX$, so:\\
We could get the optimal description from any random element: by
taking "average" of $f|_{A+x}$ over all points of the space: $x\in
X$.

\section{Statistical algorithms}

Let's say we have a statistical description, the nearer to the
optimal, the better. Now we want to construct an element using
this description. If it would be optimal - we get uniform
distribution of elements this way - we can store the same amount
of information as model's capacity.\\

We can use sequential approach like in the previous section, but
it would favor some points (e.g. the first). We would like to use
transitional invariance of the space - we cannot assume that there
were only finite number of points before.

 We are still assuming
that long range correlations vanishes, so as an approximation of
the optimal algorithm, we can assume that probability distribution
for a given point depends only on valuations of neighboring ones.

\begin{df}
Statistical algorithm \emph{is a pair } $(<,q)$:
\begin{itemize}
\item $"<" \subset X^2$ - \emph{linear order} $X$
\item $q_a:\{(x,v):v\in V(x_<)\}\to [0,1]\qquad
\left(v\cup \{(x,a)\}\notin V\Rightarrow q_a(x,v)=0\right)$
\end{itemize}
\emph{such that} $\sum_{a\in\mathcal{A}} q_a(x,v)=1$,\\
\emph{where} $x_<:=\{y\in X:y<x\}.$
\end{df}
Usually $q$ will be translational invariant and depend only on
neighboring nodes. Algorithm have to be consistent with the
model - some $q$ are enforced to 0.\\

In practice we cannot just "start" algorithm with infinite number
of previous nodes - we usually need some initialization - it will
be discussed on the end of the next section.

 While generating an element, we will get some average entropy per choice(node) - we
will count it on examples. By \emph{optimality of algorithm} we
will think: how distant is the capacity given by the algorithm to
the real capacity (model's entropy). For multidimensional models
we usually won't be able to construct optimal algorithm, only
approximate it.\\

We will analyze now two simple algorithms for Hard Square model.
\subsection{Algorithm I: Filling over independent sets}
Divide the space into separate subsets $Y_i$, inside which
constrains doesn't work (each valuation is allowed).

 For example
generally for given range of constrains $L$, take
$$Y=L\mathbb{Z}^m,\ I=\{1,...,L^m\},\
\{x^i\}_{i\in I}=\{0,...,L-1\}^m,\ \textrm{now }Y_i:=Y+x^i$$ For
HS we can take $Y_i=\{(x,y):\mathrm{mod}(x+y,2)=i\}$ - nodes
with even/odd sum of coordinates.\\

Algorithm: fill $Y_0$ using some simple probability distribution
(e.g.: with probability $q$ put 1). Now do the same with $Y_1$ (
with probability $q'$). This time only some of nodes can be valued
to 1. This second valuation doesn't influence any more nodes - we
can store here as much information as possible - take $q'=1/2$.

So our algorithm is described by one number: $q$.

Let's calculate the average entropy. Because in one half of nodes
we get entropy $h(q)$ and in the rest, 1bit/node if it's possible
- all neighbors have zeros (probability $(1-q^4$)):
 $$H_q=\frac{1}{2}h(q)+\frac{1}{2}(1-q)^4$$
The best capacity we can achieve this way is $\max_q H_q\cong
H-0.0217$\\
So we lose about 4\% of capacity: $\Delta H = 0.0217$ bit/node.\\

Why we couldn't get the optimum?\\
We can take $q$ to fulfill relation: {\small$$(p_*:=)p\left(
\begin{array}{ccc}&0&\\0&0&0\\&0&\end{array}\right)=
p\left(\begin{array}{ccc}&0&\\0&1&0\\&0&\end{array}\right)
\mathrm{,\  but\ then\ e.g.\ }
p\left(\begin{array}{ccc}1&0&\\0&0&0\\&0&\end{array}\right)>
p\left(\begin{array}{ccc}1&0&\\0&1&0\\&0&\end{array}\right)$$}\\
Explanation: If the middle node is in $Y_1$ - we have equality in
both cases($q'=1/2$). The strong inequality is got, if the middle
is in $Y_0$. Probability that we value the middle to 1 is $q$
(probability on the right side).  If we value it to $0$, we'd have
to additionally value its neighbor to
$0$, which is more probable when we fix $1$ in the corner.\\

Remind that we have countable number of equalities to fulfil
(pLOC), so usually finite number of parameters won't be enough -
we rather cannot get practical optimal algorithm, only its
approximation.
\subsection{Element generator using thermalization}
To study the next algorithm, we will need to generate valuations
on a finite set ($A$) with probability distribution close to
uniform. We could do it using approximations of statistical
algorithm. In presented here method, we will do it straightforward
- by allowing random evolution. This process will approach to kind
of thermal equilibrium - we can get to the uniform distribution as
close as we want. The disadvantage comparing to the statistical
algorithm, is that this time we have to visit every node many
times - using this method for coding is highly unpractice. But we
can use this generator for example to find the optimal statistics,
but it occurs that the approaching is slow.\\

It is based on pLOC - we will equalize iteratively
probabilities, by enforcing some random fluctuations:\\
For HS after any initialization from $V(A)$ (e.g. using some
statistical algorithm) \\
\textbf{Repeat many times: choose randomly a point in $A$. If all
of its 4 neighbors has 0 value, then change its value ($0\leftrightarrow 1$)}.\\

Let's explain why we will tend to the uniform distribution.\\
We are making some Markov process on $V(A)$. We can go from $f\in
V(A)$ to $g\in V(A)$ in one step only if they differ on exactly
one node - the probability of this transition is $\frac{1}{\#A}$
for both directions.

So the stochastic matrix describing this process is symmetric -
bistochastic - $(1,1,...,1)$ is the dominant eigenvector
corresponding to eigenvalue 1 (matrix is irreducible) - iterating
this process we will tend to the uniform distribution.

Generally we see that any bistochastic process would be
appropriate.\\

Practically I've used about 2-5 $\#A$ iteration to get the first
valuation and than to get next uncorrelated about $\#A$ iterations. \\

We can now analyze "intuitively optimal" algorithm - with random
order.
\subsection{Algorithm II: Random seed}

For every node take with uniform distribution a random real number
from [0,1], $$t:X\to [0,1]$$
It defines our order: $x<y\equiv t(x)<t(y)$\\

So in the following step of the algorithm we will take random,
unvisited node:\\
- if it has a neighbor valuated to 1 - we valuate it to $0$\\
- else - we valuate it to $0$ or $1$ with given probability.\\

We could chose this probability as a constant, but we will be more sophisticated. \\
When we are in a point with some $t$, that means that statistically
$t$ of nodes were already visited. So we can take some function:\\
\emph{charging profile} - $q:[0,1]\to [0,1]$ - if we are in a node
$t$, with probability $q(t)$ value it to 1, if we can.\\

To calculate entropy, we need to find for a given $q$ a second function:\\
$a:[0,1]\to [0,1]$ - probability that a node with given $t$ can
still be valuated to 1.

Assume that $a,q$ are continuous.\\
Now entropy will be:
\begin{equation} \label{inth} H_q=\int_0^1a(t)h(q(t))dt \end{equation} Notice that for optimality we would need:
\begin{itemize}
\item $a(0) = 1$ - we have no $1$ yet
\item $q(1)=\frac{1}{2}$ - these $1$ are not blocking anything
\item $a(1)=2p^o_*$ - they are neighbored by 4 zeros
\item $q(0)=p^o_*$ - this elements will be $1$ for sure ($a=1$)
\end{itemize}
\vbox{} Unfortunately finding $a$ from $q$ seems very difficult, so
finding optimal $q$ seems even worse.\\

\begin{figure}[h]
    \centering
        \includegraphics[width=\textwidth]{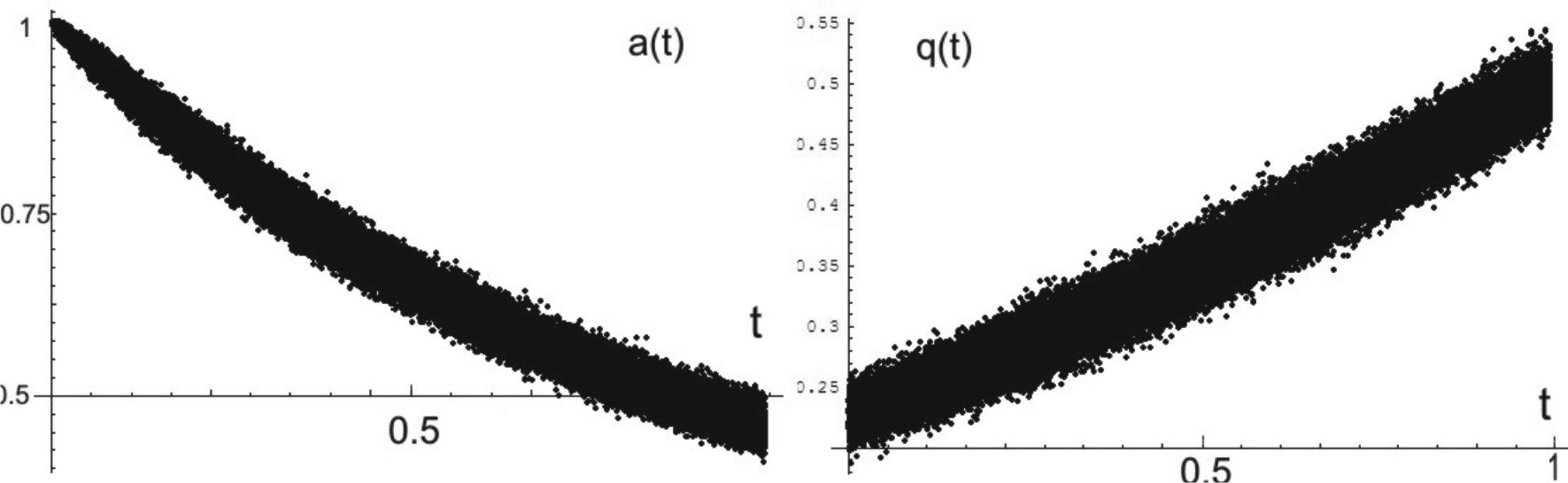}
        \caption{Numerically found $q$ i $a$.}
        \label{f2}
\end{figure}

We can approximate it numerically, using Monte-Carlo type method:\\
On some finite set (say: a square), generate different random
orders and valuations with uniform distribution (using found
generator) and take the averages to find $q$ and $a$. On
fig.\ref{f2} are shown results ($A=\{1,...,300\}^2,\ $4000
measures).

This graphs are very blurred. It's because we've simplified: $q$
parameter should depends on the order of neighboring nodes. This
algorithm looks translative invariant, but in fact only choosing
the order is so.

After fitting 4th order polynomials and integrating (\ref{inth}),
I've got entropy larger then model's. It's the lesson
that found $a$ doesn't corresponds to $q$ now.\\
We can do it exactly by generating valuations using found $q$:
I've got about $\Delta H\cong 0.01-0.02$bit/node.

\section{Approaching optimum}

In this section will be shown practical method which (if (*) is
true) can gives us description as near to the optimal one as
needed. We can use it to encode information with practically real
model's capacity. We will show numerical results for HS - there is
very good tendency to the optimum.
\subsection{Method}
The idea is to approximate the model to be able to use
one-dimensional solution:
\begin{enumerate}
\item Approximation of model - all dimensions but one (we will call it \emph{essential})
are shrunk to a finite width ($n$),
\item New alphabet - all allowed valuations of cross-section
orthogonal to the essential direction of width: range of
constrains ($L$),
\item Transfer matrix: introduce (transfer) matrix of allowed
succeeding (in essential direction) new symbols,
\item Solution to one-dimensional model - as in section 2,
\item Find algorithm - using found description. It will fill
lines succeedingly, treating every node in the same way,\\

It's good to make one more step:
\item Algorithm evaluation - use it to reconstruct the real statistical
description used and calculate entropy.
\end{enumerate}

We will go through these steps for HS:
\begin{enumerate}
\item Fix (1,0) as the essential direction on $\mathbb{Z}^2$. Fix width ($n$).\\
$$Y=\mathbb{Z}(1,0)+(1,1)\overline{n},$$ where $\overline n=\{0,...,n-1\}$.\\

We have to chose some boundary conditions. We can do it e.g. in 2
ways:
\begin{enumerate}
\item \emph{cyclic}: $\forall_i v(i+n,n)=v(i,0)$,
\item \emph{zero}: $\forall_i v(i,-1)=v(i,n)=0$.
\end{enumerate}
We can think about cyclic conditions as additional constrains, so
we get smaller entropy than original.\\
 Zero conditions - the space
is split into straps - we have all constrains instead those
between straps - lines $ni$ and $ni+1$ ($i\in\mathbb{Z}$) - we
reduce number of constrains - increase entropy.\\
 So choosing proper boundary conditions we can bound entropy from
below or above.

In this paper we are not interested in calculating entropy, but in
statistical description - we want alphabet to be as small as
possible. So the best will be cyclic conditions - thanks of
symmetries, we will be able to identify many states.
\item We are interested in valuations of $(1,1)\overline{n}$ - we
don't have constrains inside - new alphabet has $2^n$ symbols: $\mathcal{A}':=(1,1)\overline{n}\to \{0,1\}$\\
but we can identify $v, w\in\mathcal{A}'$, such that:
\begin{enumerate}
\item cyclic translation: $\exists_k\ \forall_i\
v(i)=w(i+k\  \mathrm{mod}\ n)$
\item symmetry: $\forall_i\ v(i)=w(n-1-i)$
\item unimportant zero: $v(0)=v(1)=v(2)=1\wedge w(1)=0\wedge \forall_{i\neq 1}v_i=w_i$
\end{enumerate}
We have first two conditions from the cycle symmetry.\\
The third says that two $1$ blocks neighbor of node between them -
its value is unimportant. This condition is the advantage of
taking diagonal - the number of symbols behave like $1.5^n$ and
for vertical straps: $\varphi^n\cong
1.618^n$.\\
Using this identifications for example for $n=21$ we get 3442
symbols instead of $2^{21}$ - 609 times less.
\item $M_{vw}=1\Leftrightarrow u\in V$\\
where $\mathcal{D}(u):=\{(0,0),(1,0)\}+(1,1)\overline{n},\
u(i,i):=v(i),\ u(i+1,i):=w(i)$.
\item For $M$ find the dominant eigenvector (e.g. using power method) and
the method from the second section gives the probability
distribution of two succeeding straps.
\item \textbf{Algorithm III: straps} of precision (a,b)\\
Order: $(i,j)<(k,l)\Leftrightarrow(i-j<k-l)\vee(i-j=k-l\wedge
i<k)$ \\
$q:\{0,1\}^a\times\{0,1\}^b\to[0,1]$ \\ $q(u,v)$ = if there are no neighboring $1$ and \\
$u$ - valuation of previous $a$ nodes, \\$v$ - valuation of $b$
nodes which will influence to following nodes,\\
then with this ($q$) probability valuate this node (?) to 1.\\

We get $q$ using two-straps probability distribution - look fig.
7.

\begin{figure}[h]
    \centering
        \label{qff}
        \includegraphics[width=\textwidth]{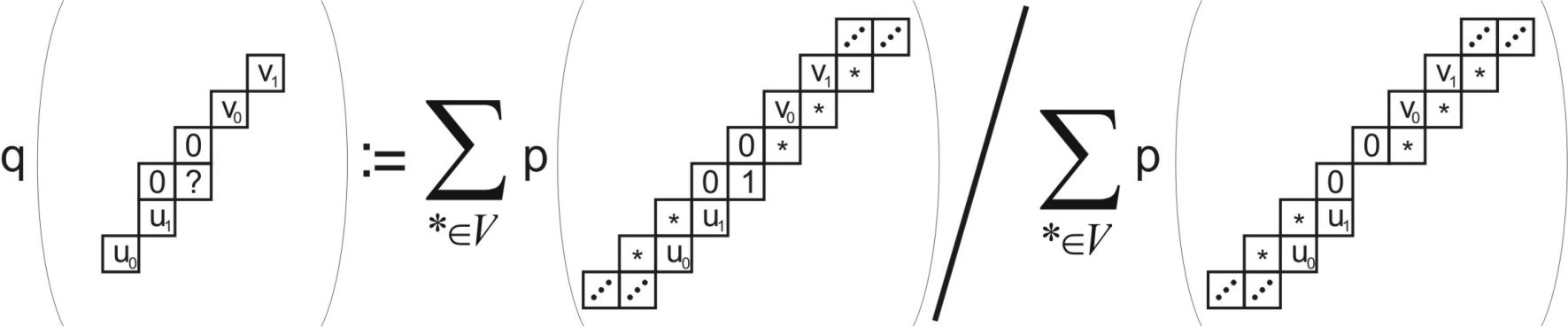}
        \caption{Finding $q$ for $a=b=2$.}
\end{figure}

Here are some algorithms as an example, found this way ($n=24$):
\begin{enumerate}
  \item $a=b=0\qquad q=0.3602994$
  \item $a=b=1\qquad
  q=\left (\begin{array}{cc}0.3365899&0.2946553\\0.4406678&0.3999231
  \end{array}\right)$
  \item $a=b=2 \qquad
  q=\left
  (\begin{array}{cccc}0.3350090&0.2918920&0.3265314&0.2910261\\
  0.3507870&0.3075943&0.3423356&0.3067152\\
  0.4419816&0.4001011&0.4342912&0.3992940\\
  0.4421921&0.4004149&0.4345211&0.3996084
  \end{array}\right)$
\end{enumerate}

Where the number of line denotes $u$ (00?,10?,01?,11?), column - $v$
analogically.
\item Algorithm evaluation.\\
We need to find statistical the description that is really
archived using given algorithm. It should be "similar" to used to
find the algorithm(we will use it as the starting point), but
while constructing the algorithm, we've lost some information. The
description we are looking for, should be fixed point of iteration
below (fig.\ref{ite}):

\begin{figure}[h]
    \centering
        \includegraphics{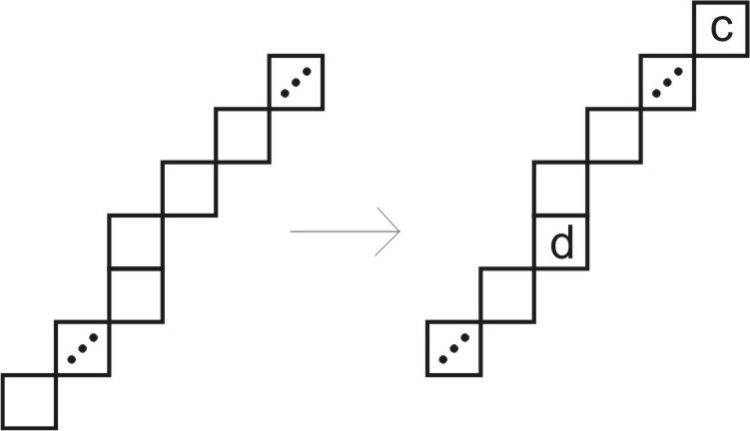}
        \caption{Iteration transforming algorithm into description}
        \label{ite}
\end{figure}

Where the probability distribution for $d$ is given by statistical algorithm.\\
There is a problem with $c$ - we cannot find it straightforward.
Although we can find its distribution, assuming that we've already
found good approximation of probability distribution of strap valuation.\\

Algorithm:
\begin{enumerate}
\item as starting description take used to find algorithm
\item while in following iteration we get description more distant
than some fixed boundary:
\begin{enumerate}
  \item using actual description find probability distribution for $(1,1)\overline{k}$
  \item using algorithm and this distribution make some iterations from the
  fig.\ref{ite}
\end{enumerate}
\end{enumerate}

While having statistical description we can count capacity we get
this way. Then we can compare it to the result from \cite{bax1} - 43
digits of model's entropy.\\

Here are results: $-\log_{10}(H-H_q)$ for different $a$, $b$:

\begin{tabular}{c|c|c|c|c|c|c|}
  $a\ \bs\ b$ & 0 & 1 & 2 & 3 & 4 & 5 \\
  \hline 0 & 2.06 & 2.113 & 2.1153 & 2.1153 & 2.1153 & 2.1153 \\
 \hline  1&3.32&3.82&3.99&4.001&4.002&4.003\\
  \hline 2&3.50&4.37&5.19&5.44&5.466&5.436\\
  \hline 3&3.50&4.42&5.55&6.43&6.74&6.78\\
  \hline 4&3.50&4.42&5.58&6.71&7.60&7.96\\
  \hline 5&3.50&4.42&5.58&6.74&7.83&8.72\\
  \hline
 \end{tabular}\\
\end{enumerate}
\subsection{Initialization}
To use found algorithm in practice, we still need to initiate it.
We could valuate the first strap anyhow and in a few straps we
would tend to assumed statistical description - we loose only some
information on the boundary.

But assume we would like to use the whole space optimally.\\
On the first look - to valuate the first strap we should use the
probability distribution for one strap. But on one side of this
strap there will be not constrains - we should be able to store
here a bit more of information.

So for the first strap, we should use the statistical description
for straps following strap filled with zeros (or suitable boundary
conditions): $p(v)=S_{0v}$. \\
Second: Straps (first($v$),second($u$)) should have probability
distribution: $p(v,u)=S_{0v}S_{vu}$ - we can find statistical
algorithm for second line from this distribution. \\
And so on. Of course we are tending to original algorithm this way.

\section{Conclusions}
\begin{itemize}
\item For one-dimensional codings, we can analytically find the optimal
statistical description - we can encode it with full capacity,
\item We have simpler alternative for arithmetic coding, which can
be used to quickly compress, encrypt and add redundancy for
correction in the same time,
\item We have criteria to ensure that given model has average
informational capacity,
\item For models in which long range correlations
vanishes, we can speak about optimal statistical description,
which gives us uniform distribution over elements,
\item If we don't need exact optimality, we can use some simple
algorithms, like filling over independent sets or over random
sequence,
\item We can generate approximations of uniform distribution on finite
set,
\item Sometimes we can find algorithm as close to optimal as needed - there are
results for Hard Square model in this paper.
\end{itemize}

\end{document}